%% file: ms.tex
\documentclass[11pt,conference,onecolumn,oneside,draftclsnofoot]{IEEEtran}
\pdfoutput=1 
\usepackage[utf8]{inputenc} 
\usepackage[T1]{fontenc} 
\usepackage{amsmath}
\usepackage{amsfonts}
\usepackage{amsbsy}
\usepackage{amssymb}
\usepackage{bbm}
\usepackage{todonotes}
\usepackage[italicdiff]{physics}
\usepackage{proba}
\usepackage{enumerate}
\usepackage{wrapfig}
\usepackage{siunitx}
\usepackage{soul} 
\usepackage{times}
\usepackage{adjustbox}
\usepackage{xfrac}
\usepackage[noadjust]{cite} 
\usepackage{url}
\usepackage{graphicx}
\usepackage{caption}
\usepackage{subcaption}
\usepackage{cuted}
\usepackage{arydshln}
\usepackage{mathrsfs}
\usepackage[tableposition=top]{caption}

\usepackage{multirow}
\usepackage{listings,multicol}    

\title{\LARGE \bf Phase-Locked Loop based Resonant Sensors:
  \\A Rigorous Theory and General Analysis Framework for\\
  Deciphering Fundamental Sensitivity Limitations due to Noise}

\author{
\IEEEauthorblockN{Alper Demir}
\vspace{0.25cm}
\IEEEauthorblockA{Ko\c{c} University, Istanbul, Turkey\\ \vspace{0.35cm}
{\small \textsf{aldemir@ku.edu.tr}}\vspace{0.0cm}}
\and
\IEEEauthorblockN{M.~Selim Hanay}
\vspace{0.25cm}
\IEEEauthorblockA{Bilkent University, Ankara, Turkey\\ \vspace{0.35cm}
{\small \textsf{selimhanay@bilkent.edu.tr}}\vspace{0.0cm}}
}

\begin{document}

\maketitle
\flushbottom

\input{abstract}
\input{introduction}

\input{pllcomponents}

\input{theory}

\input{simulation}

\input{conclusion}

\bibliographystyle{IEEEtran}
\bibliography{references}
\appendices
\input{noisecalibration}

\input{spectrumcyclo}

\input{allandev}

\end{document}

%% file: abstract.tex
\begin{abstract}
  \boldmath{} \noindent Nanomechanical resonators are used in building
  ultra-sensitive mass and force sensors. In a widely used resonator
  based sensing paradigm, each modal resonance frequency is tracked
  with a {\em phase-locked loop (PLL)} based system. There is great
  interest in deciphering the fundamental sensitivity limitations due
  to inherent noise and fluctuations in PLL based resonant sensors to
  improve their performance. In this paper, we present a precise,
  first-principles based theory for the analysis of PLL based
  resonator tracking systems. Based on this theory, we develop a
  general, rigorously-derived noise analysis framework for {PLL} based
  sensors. We apply this framework to a setting where the sensor
  performance is mainly limited by the thermomechanical noise of the
  nanomechanical resonator. The results that are deduced through our
  analysis framework are in complete agreement with the ones we obtain
  from extensive, carefully run stochastic simulations of a {PLL}
  based sensor system. We compare the conclusions we derive with the
  recent results in the literature. Our theory and analysis framework
  can be used in assessing PLL based sensor performance with other
  sources of noise, e.g., from the electronic components, actuation
  and sensing mechanisms, and due to the signal generator, as
  well as for a variety of PLL based sensor configurations such as
  multi-mode and nonlinear sensing.
\end{abstract}

%% file: introduction.tex
\section{Introduction}
\label{sec:intro}
\noindent
State-of-the-art nano-mechanical sensors are extremely sensitive,
achieving yoctogram and single-protein resolutions in inertial mass
sensing, thanks to their ever diminishing size and high quality
factors~\cite{chaste2012nanomechanical,naik2009towards,hanay2012single}. Currently,
there are two main architectures in use for resonant sensors: (i)
Self-sustaining autonomous oscillator, in which a nano-mechanical
resonator is used as the frequency selective element in a classic
feedback oscillator configuration, with an amplifier and a delay line
in the loop~\cite{feng2008self,van2013nonlinear}. (ii) Phase-locked
loop ({\sc {\sc PLL}}) configuration that tracks the nano-mechanical
resonance frequency by driving the resonator with a
voltage/numerically-controlled oscillator which is locked to the
resonance. Both architectures have their advantages and
shortcomings. However, the accuracy of all {\sc NEMS} sensors is
limited by the inherent fluctuations and noise in the mechanical,
electrical and/or optical
domains~\cite{cleland2002noise,ekinci2004ultimate,gavartin2013stabilization,Sansa2016},
depending on the particular sensing and actuation mechanisms used.
Thus, there is great interest in first understanding the fundamental
sensitivity limitations due to noise, and then improving the sensor
performance~\cite{cleland2002noise,ekinci2004ultimate,gavartin2013stabilization,Sansa2016,kenig2012optimal,villanueva2013surpassing,kenig2013phase,demir2018numerical,roy2018improving}.

Recently, Roy~{\em et~al.}~\cite{roy2018improving} put forward an idea
that is diametrically opposed to the current understanding on how to
improve NEMS sensor performance. They proposed that one can achieve
much better sensitivities by increasing damping in the resonator,
i.e., with resonators that have much lower quality factors. They offer
a two-part argument as to how this improvement can be obtained. In the
first part, they point out that nano-mechanical resonators with lower
quality factors can be driven harder before Duffing nonlinearity kicks
in. Higher drive strength in conjunction with larger damping turns the
inherent thermo-mechanical noise of the resonator into the dominant
source of noise, masking other sources of noise, and enables operation
at a higher signal-to-noise-ratio
($\text{\small\textsf{SNR}}$). Roy~{\em et~al.} adjust the drive
strength in such a way so that $\text{\small\textsf{SNR}}$ is
inversely proportional to the quality factor $Q$, thus operating at
the onset of Duffing nonlinearity. This first part of their proposal
makes perfect sense. In the second part, Roy~{\em et~al.} claim that,
with $\text{\small\textsf{SNR}}\propto 1/Q$, one in fact obtains a
noise performance that is much better than the expected with a {\sc
  PLL} based architecture at a lower $Q$. Their explanation regarding
as to how this improvement arises for low $Q$ is based on a revelation
that the phase noise spectrum flattens at low frequencies, as opposed
to the usual approximation that is commonly employed in high $Q$
cases.  We believe that this second part of their claim warrants
further investigation.

\begin{figure}[t]
\centering
\includegraphics[width=0.49\textwidth]{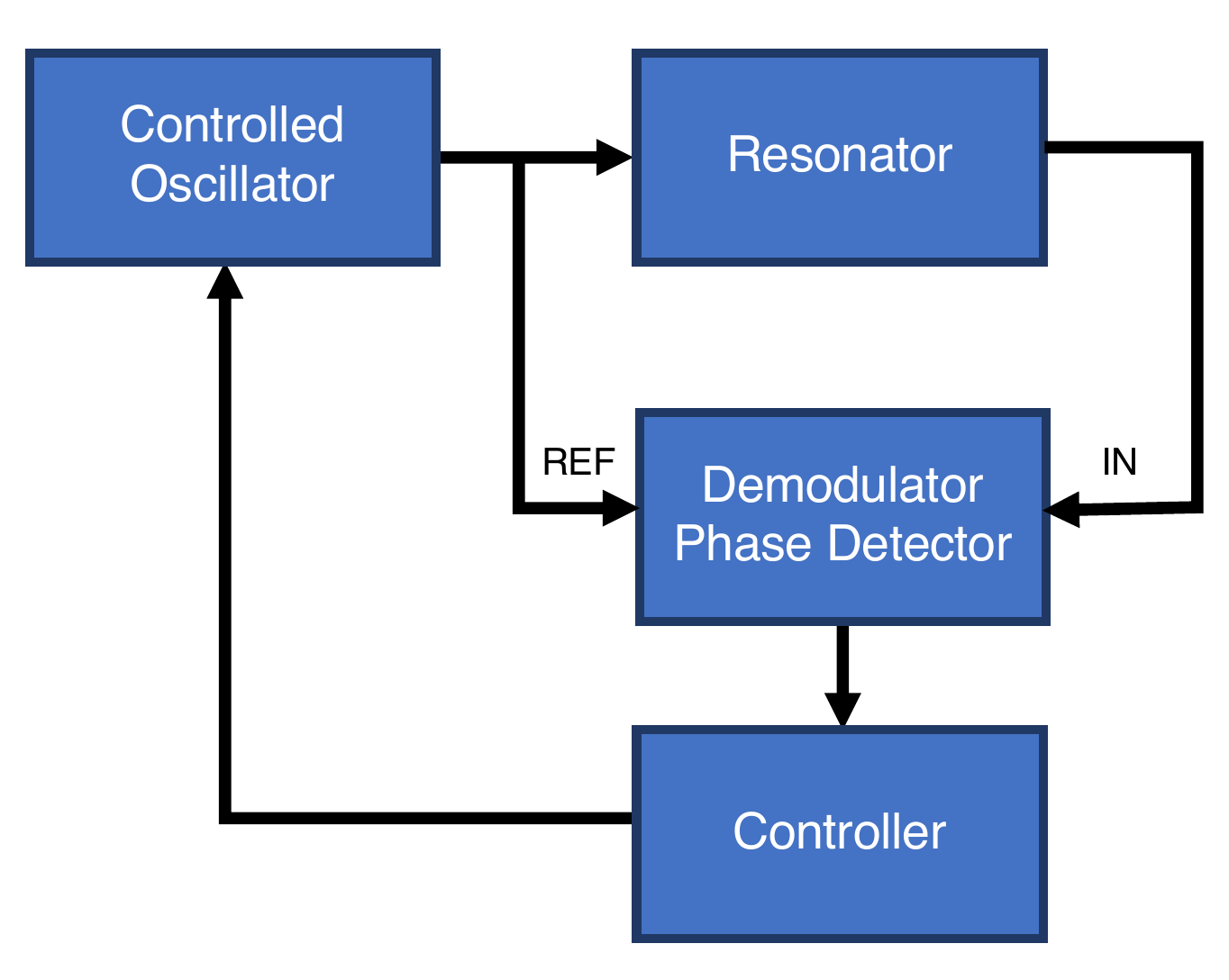}
\caption{Phase-locked loop based tracking of a resonator}
\label{fig:plltracking}
\end{figure}

\begin{figure*}[t]
\centering
\includegraphics[width=0.99\textwidth]{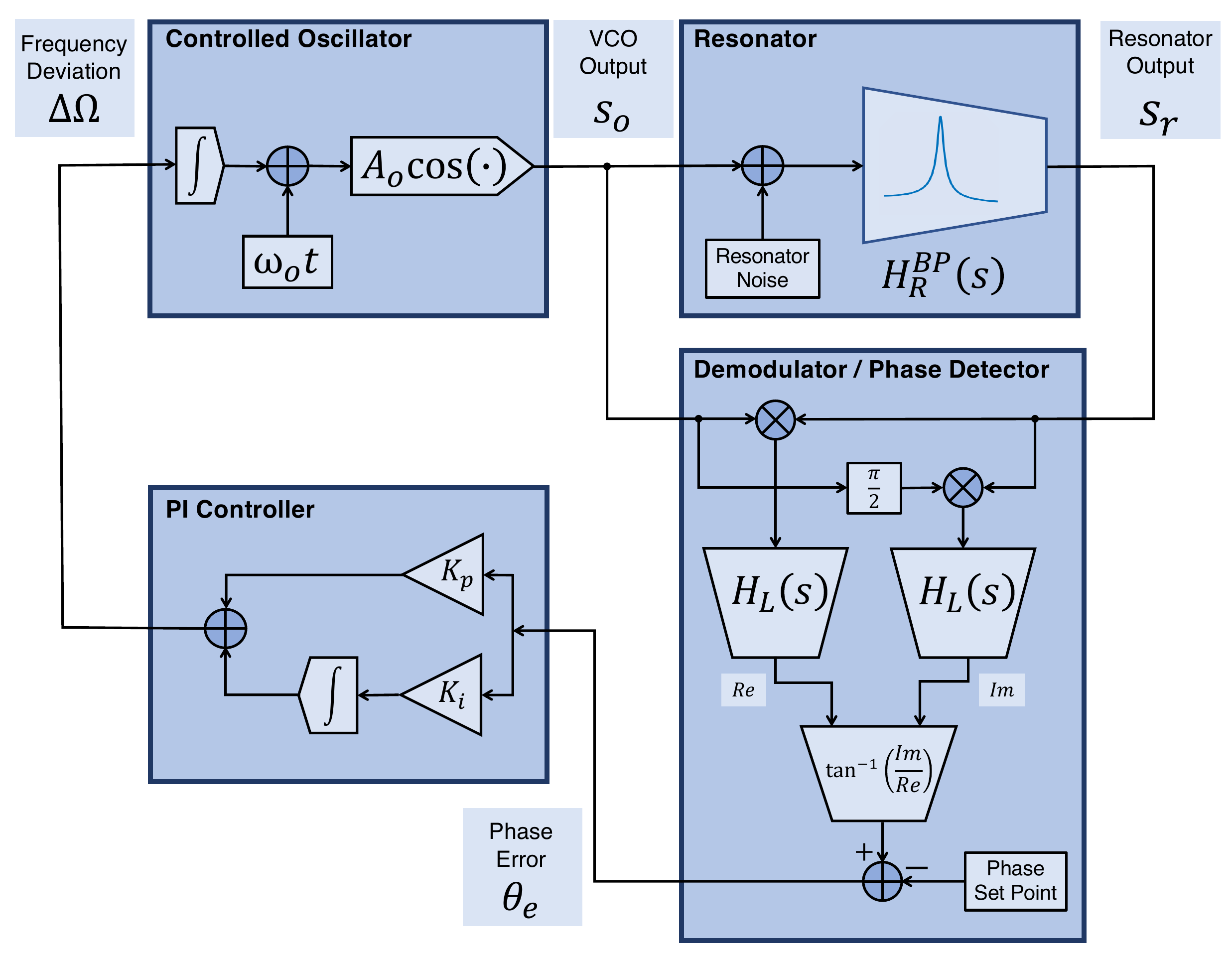}
\caption{Full model of PLL based resonator tracking}
\label{fig:pllfull}
\end{figure*}

In this paper, we get to the bottom of this issue. 
We investigate in detail whether one can obtain better performance
with larger damping in a PLL based sensor. Our conclusion is in the
negative. We arrive at this result by developing a first-principles
based noise analysis framework for {\sc PLL} based sensor
architectures, where each step and approximation is rigorously
justified. The {\sc PLL} system we consider is shown in
Figure~\ref{fig:plltracking} as a block diagram, with details shown in
Figure~\ref{fig:pllfull}.  We apply our analysis framework to the case
considered by Roy~{\em et~al.}. We precisely characterize the
performance of a {\sc PLL} based sensor when the dominant source of
noise is the inherent thermo-mechanical noise of the resonator.  We
perform the noise analysis for different $Q$ values and feedback
parameters.  Moreover, our analysis framework is much more general. It
can be used to assess the performance of {\sc PLL} based {\sc NEMS}
sensors with other sources of noise, e.g., from the amplifiers,
actuation and sensing mechanisms, and due to the signal generator, as
well as for a variety of {\sc PLL} based sensor configurations such as
multi-mode~\cite{hanay2012single} and nonlinear
nano-mechanical~\cite{yuksel2019nonlinear} sensing.
Furthermore, the theory we develop is not specific to nanomechanical
resonators, it may be used for resonant sensors in other domains, for
instance ones that are based on microwave resonators~\cite{kelleci2018towards}. 

Our analysis framework does employ some approximations, and is based
on some assumptions. Even though all of these are stated very clearly
and justified rigorously in our development, we go further in order to
verify our theoretical results.  We report the results of extensive,
carefully conceived stochastic simulations of the {\sc PLL}
system. {\bf The simulator does not employ any of the mentioned
  approximations and assumptions, and is based on realistic, detailed,
  full models of the system components.}  The system is simulated with
full, high-frequency, nonlinear and time-varying models for the
resonator and the demodulator as shown in Figure~\ref{fig:pllfull}.
The simulator is based on the solution of coupled
differential equations that are solved using an appropriate numerical
technique based on time-discretization. The time step of the
simulation is set to be a small fraction ($\approx 1/100$) of the
period of the high-frequency signal at the output of the
resonator. The thermomechanical noise of the resonator is introduced
into the simulation using a random number generator. The time series
and waveform data produced by the simulation is post-processed for estimating the
spectral densities of the signals of interest, as well as in order to
compute the Allan Deviation for the frequency tracking performance of
the closed-loop system. The results we obtain with the simulator are
in complete agreement with the ones deduced via our analysis
framework.  

The outline of the paper is as follows. We present the models used for
the {\sc PLL} components in Section~\ref{sec:pllcomponents}. The theoretical
development of our analysis framework is in Section~\ref{sec:theory}.
The results obtained from our theory are verified against the ones
obtained with the simulator in
Section~\ref{sec:simulation}. Conclusions are stated in
Section~\ref{sec:conc}. Three appendices provide derivations and details,
regarding thermo-mechanical resonator noise, spectral characterization
and filtering of cyclo-stationary random processes, and Allan Variance, in order
to make the paper as self-contained as possible.

%% file: pllcomponents.tex
\section{PLL Component Models}
\label{sec:pllcomponents}
\subsection{Resonator} 
We consider a resonator that is modeled as a damped harmonic
oscillator as follows~\cite{hauer2013general} 
\begin{equation}
\dv[2]{}{t}\,x + \Gamma\,\dv{}{t}\,x + \omega_r^2\,x = \frac{F\pqty{t}}{m}
\end{equation}
where $x$ is the displacement, $m$ is the mass, $F\pqty{t}$ represents a
force excitation, $\omega_r$ is the
resonance frequency, and the damping rate $\Gamma$ is given by 
\begin{equation}
\Gamma = \frac{\omega_r}{Q} 
\label{eqn:gammadefn}
\end{equation}
where $Q$ is the quality factor. The damping rate $\Gamma$
determines the line-width of the resonator's frequency response (from
input $F\pqty{t}$ to output $x\pqty{t}$), which is given by
\begin{equation}
H_{R}^{BP}\pqty{s}=\frac{X\pqty{s}}{F\pqty{s}}=\frac{1}{m}\,\frac{1}{s^2+\Gamma\,s+\omega_r^2}
\label{eqn:resonatortranfun}
\end{equation}
Based on the {\it fluctuation–dissipation theorem} of statistical
thermodynamics, the {\it thermo-mechanical noise} of the resonator can be
modeled as a white noise source (input-referred, at the input of the
resonator as in Figure~\ref{fig:pllfull}) with a (two-sided) spectral
density given by~\cite{ekinci2004ultimate}
\begin{equation}
S_{thm}\pqty{\omega} = 2\,m\,\Gamma\,k_{B}\,T
\label{eqn:thmnoisePSD} 
\end{equation}
where $k_B$ is Boltzmann's constant, and $T$ is temperature (in
Kelvins). With this noise source as the only input to the resonator,
the mean kinetic energy of the resonator can be computed as follows
\begin{equation}
E_{K} = \vb{E}\bqty{\frac{1}{2}m\, \pqty{\dv{}{t}\,x}^2 } 
\end{equation}
where $ \vb{E}\bqty{\cdot}$ denotes the probabilistic expectation
operator. The expectation above can be computed using the
{\it Wiener–Khinchin theorem} as below
\begin{equation}
E_{K} = \frac{1}{2}m\,\frac{1}{2\pi} \int_{-\infty}^{\infty}
\omega^2\;\vb{S}_x\pqty{\omega}\,\dd \omega
\label{eqn:thmnoiseintegral} 
\end{equation} 
where $\vb{S}_x\pqty{\omega}$ is the power spectral density ({\sc
  PSD}) of the resonator displacement $x(t)$, which can be computed
with
\begin{equation}
\vb{S}_x\pqty{\omega} = \vqty{H_{R}^{BP}\pqty{j\,\omega}}^2\;S_{thm}\pqty{\omega}
\end{equation} 
where $j=\sqrt{-1}$. The integral in \eqref{eqn:thmnoiseintegral}
can be evaluated, as shown in Appendix~\ref{sec:thermechnoiseintegral},
to yield 
\begin{equation}
E_{K} = \frac{k_{B}\,T}{2}
\end{equation} 
consistent with the {\it equipartition theorem} of statistical mechanics. 
\subsection{Demodulator} 
\label{sec:demodulator}
The demodulator shown in Figure~\ref{fig:pllfull} performs essentially
as a {\em phase (difference) detector}. The controlled oscillator output
and the resonator output can be expressed as 
\begin{equation}
\begin{aligned}
s_o\pqty{t} & = & A_o\,\cos\pqty{\omega_o\,t+\theta_o\pqty{t}}\\
s_r\pqty{t} & = & A_r\,\cos\pqty{\omega_r\,t+\theta_r\pqty{t}}
\end{aligned}
\end{equation}  
We can express the operations in the in-phase (real) and quadrature
(imaginary) arms of the demodulator in a compact manner
using complex arithmetic as follows. We first express the resonator
output as 
\begin{equation}
s_r\pqty{t} =  \frac{A_r}{2}\,\bqty{e^{j\,\pqty{\omega_r\,t+\theta_r\pqty{t}}}+e^{-j\,\pqty{\omega_r\,t+\theta_r\pqty{t}}}}
\end{equation} 
Then, the signals at the output(s) of the multipliers in the
demodulator are the real and imaginary parts of  
\begin{equation}
\begin{aligned}
\frac{A_r A_o}{2}&\bqty{e^{j\pqty{\omega_rt+\theta_r\pqty{t}}}+e^{-j\pqty{\omega_rt+\theta_r\pqty{t}}}}e^{-j\pqty{\omega_ot+\theta_o\pqty{t}}}\\
 = \frac{A_r A_o}{2} &\left [e^{j\pqty{\pqty{\omega_r-\omega_o}t+{\theta_r\pqty{t}-\theta_o\pqty{t}}}}\right.+\\
& \quad\quad\quad\quad\quad\quad\left. e^{-j\pqty{\pqty{\omega_r+\omega_o}t+{\theta_r\pqty{t}+\theta_o\pqty{t}}}}\right]
\end{aligned}
\label{eqn:demodsignal}
\end{equation} 
We next assume that the low-pass filters in the demodulator (shown as
$H_L\pqty{s}$) block the high-frequency, second term and pass the
low-frequency, first term above. Thus, the outputs of the filters are
given by (the real/imaginary parts of) 
\begin{equation}
\frac{A_r A_o}{2}\;e^{j\pqty{\pqty{\omega_r-\omega_o}t+{\theta_r\pqty{t}-\theta_o\pqty{t}}}}
\label{eqn:demodoutdet}
\end{equation} 
Finally, the output of the $\arctan\pqty{\cdot}$ block is then
simply the phase difference between the resonator output and the
control oscillator signal, expressed as 
\begin{equation}
\theta_e\pqty{t} = \pqty{\omega_r-\omega_o}\,t+ \theta_r\pqty{t}-\theta_o\pqty{t}
\end{equation} 
where we assume that $\omega_r\approx\omega_o$ and
$\theta_r\pqty{t}-\theta_o\pqty{t}$ is a low-frequency signal.  
In the actual model of the system, we do take into account the
non-ideal nature of the low-pass filter $H_L\pqty{s}$ by using a
practical filter. The attenuation
of the high-frequency component in \eqref{eqn:demodsignal} by a
practical $H_L\pqty{s}$ is in fact quite good due to the large
frequency separation, whereas the low-frequency component does get
modified by the filter, which we take into account in the analytical
theory we develop further below. The phase set point 
in the demodulator is needed in order to keep the resonator at its
resonance, as we show later. 
\subsection{Controller} 
We use a simple {\sc PI} controller, as shown in
Figure~\ref{fig:pllfull}, with a transfer function
\begin{equation}
H_{PI}\pqty{s} = K_p + \frac{K_i}{s}
\label{eqn:picontrollertransfun}
\end{equation} 
We discuss later how to choose the controller parameters $K_p$ and
$K_i$. The input to the controller is the phase error signal from the
demodulator (phase difference detector) and the controller output is
fed to the controlled oscillator, simply determining its frequency
deviation from the nominal value $\omega_o$.
\subsection{Controlled  oscillator} 
The controlled oscillator is an essential component of the {\sc
  PLL}. In an all-analog {\sc PLL} system, it would be instantiated as
a high precision analog {\it voltage-controlled oscillator} ({\sc
  VCO}), typically including a crystal as a time reference. In {\sc PLL}
tracking systems for {\sc NEMS} applications, {\it lock-in amplifiers}
({\sc LIA}) are routinely employed. In recent {\sc LIA} based systems,
the controlled oscillator is in fact digitally implemented on a
configurable {\sc DSP/FPGA} chip, as what is called a {\it numerically
  controlled oscillator} ({\sc NCO}). The timing/frequency precision
of the {\sc NCO} is then determined by the time base of the {\sc
  DSP/FPGA}, which may be locked to an external atomic reference. For
the {\sc VCO/NCO}, we will use a simple model as follows for its
output
\begin{equation}
s_o\pqty{t} =
A_o\cos\pqty{\omega_o\,t+\int^{t}\Delta\Omega\pqty{\tau}\,\dd\tau}
\end{equation}   
where the frequency deviation $\Delta\Omega\pqty{t}$ is the control
signal produced by the {\sc PI} controller.

Due to the high precision of the {\sc VCO} or the time base of the
{\sc NCO}, we will assume that the phase noise contribution of the
controlled oscillator is negligible. However, both the theory and the
simulator that we have developed can be very easily modified to
include the phase noise contribution from the VCO/NCO, as well as
noise from other components, such as the electronic amplifiers,
the actuators and sensors that convert signals between the electrical and
the mechanical domain. Our main goal in this paper is to decipher the
fundamental sensitivity limitation due to resonator noise.

%% file: theory.tex
\section{Theory}
\label{sec:theory}
\subsection{Baseband equivalent phase domain model of the resonator}
The {\sc PLL} based tracking system shown in Figure~\ref{fig:pllfull}
contains signals with widely varying frequencies, as well as in
different {\it domains}. The outputs of the {\sc NCO} and the
resonator are at a high frequency, at around the resonance frequency
of the resonator. On the other hand, the phase error signal at the
output of the phase detector and the frequency deviation at the output
of the controller are low-frequency signals, at frequencies below the
{\sc PLL} bandwidth. The frequency separation between these high and
low frequency signals is typically at least four orders of
magnitude. The signal of interest is the frequency deviation at the
input of the {\sc NCO}, since that is what one uses in a typical
sensing setup in order to track the resonance frequency
deviations. The challenge in analyzing the {\sc PLL} system is then to
accurately characterize the slow dynamics of the frequency deviation
signal while capturing the impact of the fast dynamics of the {\sc
  NCO} and the resonator in a correct manner. In order to accomplish
this in a simple and tractable manner, we will first develop a
base-band (low-frequency) equivalent model of the
NCO-resonator-demodulator signal chain. The input and output of this
chain of blocks are both low-frequency signals, whereas there is first
low-to-high and then high-to-low {\it frequency translation} of
signals along the chain, and also a nonlinear $\arctan\pqty{\cdot}$
operation at the very end. This makes this composite system, the
cascade connection of {\sc NCO}-resonator-demodulator, both {\it
  nonlinear} and {\it time-varying}. Fortunately, we will be able to
develop a simple, {\it linear} and {\it time-invariant} model for this
cascade that is quite accurate, which is verified against numerical
simulations of the full, nonlinear and time-varying system.

\begin{figure*}[t]
\centering
\includegraphics[width=0.99\textwidth]{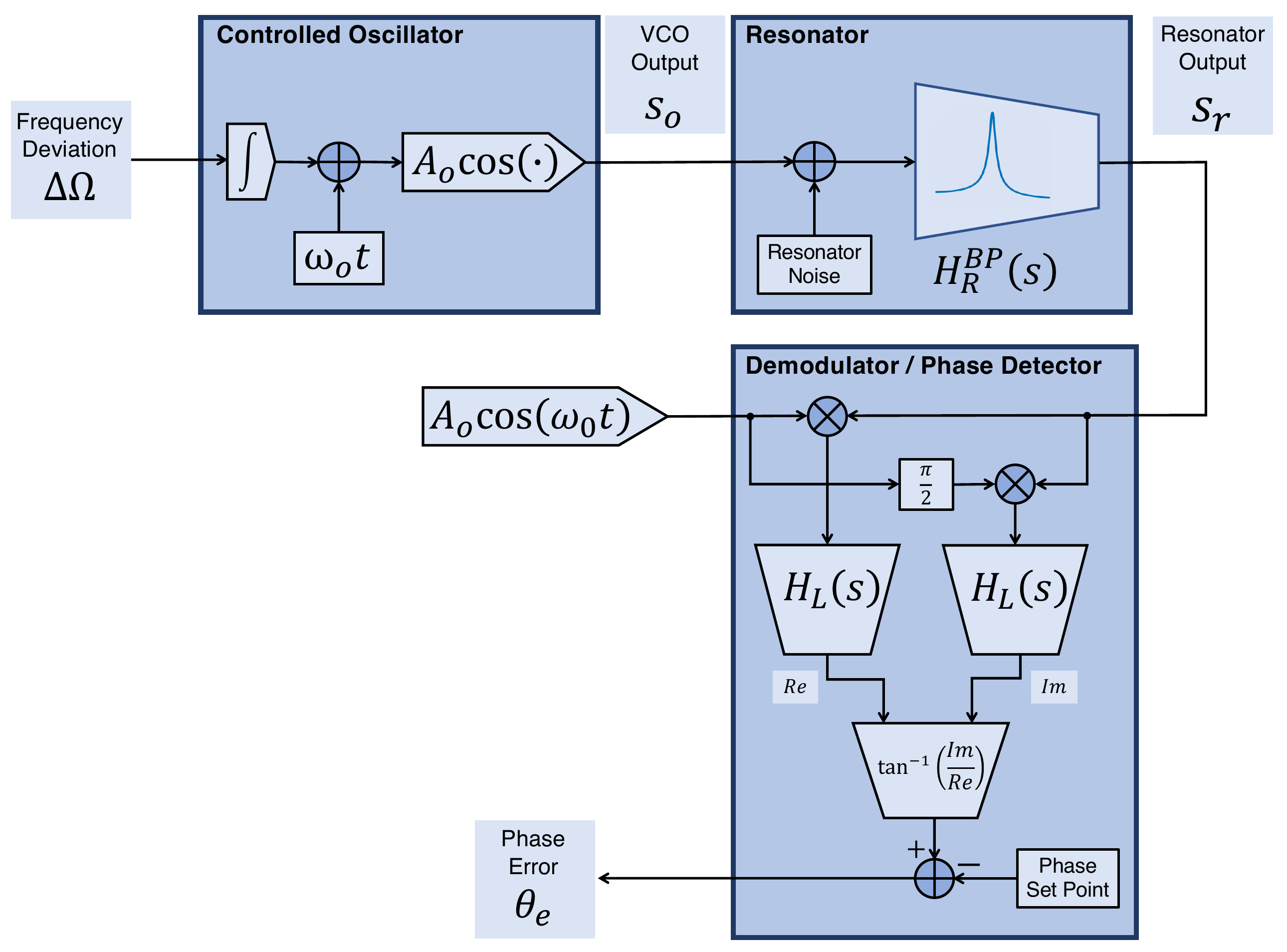}
\caption{Open-loop controlled oscillator-resonator-demodulator chain}
\label{fig:pllopen}
\end{figure*}

In order to develop this simple model, we consider the open-loop {\sc
  NCO}-resonator-demodulator chain shown in
Figure~\ref{fig:pllopen}. We note that, in constructing the open-loop
model, not only we have disconnected the main {\sc PLL} loop but also
we no longer feed the second demodulator input with the signal from
the {\sc NCO}. Instead, the second demodulator input is simply set to
a sinusoidal signal at a constant frequency. In the final simplified
model we will construct, we will take into account the fact that the
second demodulator input is in fact set to the {\sc NCO} output. In
the development below, we set the resonator noise to zero and first
construct a simplified model for the deterministic dynamics. We will
then also consider the resonator noise and include it in the final
model.

We now walk through the chain, from the input (frequency deviation)
$\Delta\Omega\pqty{t}$ to the output (phase error) $\theta_e\pqty{t}$. 
The phase deviation of the {\sc NCO}, $\theta_o\pqty{t}$ is related to the
frequency deviation simply with an integral
\begin{equation}
\theta_o\pqty{t} = \int^{t}\Delta\Omega\pqty{\tau}\,\dd\tau
\label{eqn:freqtophase}
\end{equation} 
The output of the {\sc NCO} is then given by 
\begin{equation}
\begin{aligned}
s_o\pqty{t} & = A_o\cos\pqty{\omega_o\,t+\theta_o\pqty{t}}\\
& =  \frac{A_o}{2}\bqty{e^{j\,\pqty{\omega_o\,t+\theta_o\pqty{t}}}+e^{-j\,\pqty{\omega_o\,t+\theta_o\pqty{t}}}}
\end{aligned}
\label{eqn:NCOsignal} 
\end{equation} 
The operation above constitutes a {\it low-to-high frequency translation},
from $\theta_o\pqty{t}$ to $s_o\pqty{t}$. 
We next consider only the first term on the second line of
\eqref{eqn:NCOsignal}. As discussed in Section~\ref{sec:demodulator},
the signal that will arise from the second term down the
resonator-demodulator signal chain will be eventually blocked by the
low-pass filter $H_L\pqty{s}$. Then, the input to the
resonator is given by 
\begin{equation}
s_{inr}\pqty{t} =  \frac{A_o}{2}\,
e^{j\,\pqty{\omega_o\,t+\theta_o\pqty{t}}} =
\frac{A_o}{2}\,e^{j\omega_o t}\:e^{j\theta_o\pqty{t}}
\label{eqn:resonatorinput} 
\end{equation} 
In order to compute the effect of the resonator on the signal, we now
transform to frequency domain, by computing the Laplace transform
(bilateral) of $s_{inr}\pqty{t}$ above:
\begin{equation}
S_{inr}\pqty{s} =  {\cal L}\Bqty{\frac{A_o}{2}\,e^{j\omega_o
    t}\:e^{j\theta_o\pqty{t}}} = S_{inr}^{BB}\pqty{s-j\omega_o}
\label{eqn:resonatorinputlaplace} 
\end{equation} 
where $S_{inr}^{BB}\pqty{s}$ is 
\begin{equation}
S_{inr}^{BB}\pqty{s} =  {\cal
  L}\Bqty{\frac{A_o}{2}\:e^{j\theta_o\pqty{t}}} = {\cal
  L}\Bqty{s_{inr}^{BB}\pqty{t}} 
\label{eqn:resonatorinputlaplaceBB} 
\end{equation} 
While $s_{inr}\pqty{t}$ is a high-frequency, pass-band signal with its
power concentrated around $\omega_o$ in the frequency domain,
$s_{inr}^{BB}\pqty{t}$ is a base-band, low-frequency signal with its power
concentrated around zero frequency. When the {\sc PLL} is tracking the
resonance of the resonator, the {\sc NCO} center frequency $\omega_o$
would be nominally equal to the resonant frequency $\omega_r$ of the
resonator. If there is any resonance frequency shift in the resonator,
{\sc PLL} would compensate for this by adjusting the frequency
deviation $\Delta\Omega\pqty{t}$, and accordingly the phase deviation 
$\theta_o\pqty{t}$, of the {\sc NCO}. Thus, without loss of
generality, we assume $\omega_o = \omega_r$ as we proceed below. 

We now compute the output of the resonator (in the frequency domain)
with its input as in \eqref{eqn:resonatorinputlaplace}
\begin{equation}
S_{outr}\pqty{s} = H_{R}^{BP}\pqty{s}\:S_{inr}\pqty{s}  
\label{eqn:resonatoroutlaplace} 
\end{equation} 
where $H_{R}^{BP}\pqty{s}$ is the resonator transfer function in
\eqref{eqn:resonatortranfun}. Nominally, $H_{R}^{BP}\pqty{s}$ has a
pass-band characteristics centered around $\omega_o$ in the frequency
domain. Thus, with the input $s_{inr}\pqty{t}$ as a pass-band signal
centered around the same frequency, so is the output
$s_{outr}\pqty{t}$. Hence, we have
\begin{equation}
s_{outr}\pqty{t} = e^{j\omega_o t}\:s_{outr}^{BB}\pqty{t}
\label{eqn:resonatorout}
\end{equation}
and 
\begin{equation}
S_{outr}\pqty{s} = S_{outr}^{BB}\pqty{s-j\omega_o}
\label{eqn:resonatoroutlaplaceBB}
\end{equation}
where $s_{outr}^{BB}\pqty{t}$ is a base-band, low-frequency signal with its power
concentrated around zero frequency. Combining
\eqref{eqn:resonatorinputlaplace}, \eqref{eqn:resonatoroutlaplace} and
\eqref{eqn:resonatoroutlaplaceBB}, we obtain
\begin{equation}
\begin{aligned}
S_{outr}^{BB}\pqty{s-j\omega_o} &=& H_{R}^{BP}\pqty{s}\:S_{inr}^{BB}\pqty{s-j\omega_o}\\
S_{outr}^{BB}\pqty{s} &=& H_{R}^{BP}\pqty{s+j\omega_o}\:S_{inr}^{BB}\pqty{s}
\end{aligned}
\end{equation}
We define 
\begin{equation}
H_{R}^{BB}\pqty{s}=H_{R}^{BP}\pqty{s+j\omega_o}
\label{eqn:resonatortranfunBB}
\end{equation}
as the {\it base-band equivalent transfer function of the
resonator}, which is given by 
\begin{equation}
H_{R}^{BB}\pqty{j\omega} =
\frac{1}{m}\,\frac{1}{-\pqty{\omega+\omega_o}^2+\frac{\omega_o}{Q}\,
  j\,\pqty{\omega+\omega_o}+\omega_o^2}
\label{eqn:resonatorfreqrespBB}
\end{equation}
where we used \eqref{eqn:resonatortranfun} and
\eqref{eqn:resonatortranfunBB}, and substituted $s=j\omega$ and
$\Gamma=\omega_o/Q$. $H_{R}^{BB}\pqty{j\omega}$ can be used as a
base-band equivalent model for the resonator. However, an approximate
(first-order) form for $H_{R}^{BB}\pqty{j\omega}$ (second-order) 
that we derive below simplifies the rest of our derivations considerably.  

We manipulate \eqref{eqn:resonatorfreqrespBB} (by combining the first
and third terms in the denominator of the $\omega$ dependent part) to obtain 
\begin{equation}
H_{R}^{BB}\pqty{j\omega} =
\frac{1}{m}\,\frac{1}{-\omega\,\pqty{\omega+2\,\omega_o}+ j\,\frac{\omega_o}{Q}\,
 \pqty{\omega+\omega_o}}
\label{eqn:resonatorfreqrespBB2}
\end{equation}
We note that $\omega$ in \eqref{eqn:resonatorfreqrespBB2} above is
small when compared with $\omega_o$. This is due to the frequency
shift operation represented by \eqref{eqn:resonatortranfunBB}. In the
pass-band model of the resonator represented by
$H_{R}^{BP}\pqty{j\omega}$, we have $\omega\approx \omega_o$, whereas in the
base-band equivalent model represented by
$H_{R}^{BB}\pqty{j\omega}=H_{R}^{BP}\pqty{j\pqty{\omega+\omega_o}}$, we
  have $\omega \approx 0$.    
We then assume that $\omega\ll\omega_o$ in 
\eqref{eqn:resonatorfreqrespBB2}
and use the following approximations due to low-frequency,
base-band nature of $H_{R}^{BB}\pqty{j\omega}$
\begin{equation}
\omega+2\,\omega_o \approx 2\,\omega_o\quad,\quad
\omega+\omega_o \approx \omega_o 
\label{eqn:approxforfregresp}
\end{equation}
The above can be interpreted as a sort of high-$Q$ approximation, but
our goal is to develop a theory that is valid even for low-$Q$
resonators. We verify later against simulations (which do not
incorporate any approximations) that the base-band resonator model
based on the above approximations remains accurate for a $Q$ that is
as low as 10. With the above approximations,
$H_{R}^{BB}\pqty{j\omega}$ can be simplified as follows
\begin{equation}
\begin{aligned}
H_{R}^{BB}\pqty{j\omega} 
& \approx \frac{1}{m}\,\frac{1}{-2\,\omega\,\omega_o+ j\,\frac{\omega_o^2}{Q}}\\ 
&= \frac{1}{m}\,\frac{1}{j}\frac{1}{\frac{\omega_o^2}{Q}+j\,2\,\omega\,\omega_o}\\
&=\frac{1}{m}\,e^{-j\,\frac{\pi}{2}}\frac{1}{\frac{\omega_o^2}{Q}+j\,2\,\omega\,\omega_o}\\
&=\frac{Q}{m\,\omega_o^2}\,e^{-j\,\frac{\pi}{2}}\frac{1}{1+j\,\omega\,\frac{2\,Q}{\omega_o}}\\
\end{aligned}
\label{eqn:resonatorfreqrespBB3}
\end{equation}
Finally, $H_{R}^{BB}\pqty{j\omega}$ can be represented as a Laplace
transform:
\begin{equation}
H_{R}^{BB}\pqty{s} 
=\frac{Q}{m\,\omega_o^2}\,e^{-j\frac{\pi}{2}}\frac{1}{1+s\,\frac{2\,Q}{\omega_o}}\\
\label{eqn:resonatorfreqrespBB4}
\end{equation}
We define the resonator time constant with 
\begin{equation}
\tau_{r}=\frac{2\,Q}{\omega_o} = \frac{2}{\Gamma}
\label{eqn:resonatorTC}
\end{equation}
and obtain 
\begin{equation}
H_{R}^{BB}\pqty{s} 
=\frac{Q}{m\,\omega_o^2}\,e^{-j\frac{\pi}{2}}\frac{1}{1+s\,\tau_r}\\
\label{eqn:resonatorfreqrespBB5}
\end{equation}
The above is essentially a first-order, one-pole, low-pass transfer
function, with a DC gain and an extra phase shift. If the input to the
resonator is a pure tone at the resonance frequency $\omega_o$
(corresponding to $s=0$ in \eqref{eqn:resonatorfreqrespBB5}), then the
steady-state output (also a pure tone at the same frequency) will have
a $-\pi/2$ phase shift with respect to the input. 

We note that a resonator model as in \eqref{eqn:resonatorfreqrespBB5}
was derived in~\cite{olcum2015high}. However, our treatment above
based on the use of the base-band equivalent transfer function concept
streamlines the model development process and reveals the exact nature
of the approximations involved. The base-band equivalent
representation for band-pass signals is commonly used in the analysis
of communication systems~\cite{benedetto1999principles}. This
technique is similar to the ones used in other disciplines, known as
complex amplitude representation for slow
dynamics~\cite{kenig2012optimal}, and slowly varying envelope
approximation~\cite{yurke1995theory}.

Next, we move along the signal chain and characterize the impact of
the demodulator. The demodulator features {\it high-to-low frequency
  translation}, undoing the {\it low-to-high frequency translation}
that was done by the {\sc NCO}. That is, the signals at the output(s)
of the multipliers in the demodulator, in Figure~\ref{fig:pllopen},
are the real and imaginary parts of
\begin{equation}
s_{m}\pqty{t} = A_o\,e^{-j\omega_o t}\:s_{outr}\pqty{t}
\label{eqn:demodoutput1}
\end{equation}
We substitute \eqref{eqn:resonatorout} into the above equation to
obtain 
\begin{equation}
s_{m}\pqty{t} = A_o\,e^{-j\omega_o t}\:e^{j\omega_o
  t}\:s_{outr}^{BB}\pqty{t} = A_o\,s_{outr}^{BB}\pqty{t} 
\label{eqn:demodoutput}
\end{equation}
That is, the demodulator simply extracts the base-band equivalent,
low-frequency, complex-valued resonator output
$s_{outr}^{BB}\pqty{t}$. However, this signal is further processed in
the demodulator through the low-pass filter denoted by
$H_L\pqty{s}$. This filter will nominally not modify
$s_{outr}^{BB}\pqty{t}$. However, it is required in order to remove
the high-frequency signals (at the outputs of the multipliers) that
will arise from the second term in \eqref{eqn:NCOsignal}, which we
have ignored upfront. 

The phase angle of the complex-valued output of the low-pass
filters is produced with the $\arctan\pqty{\cdot}$ block in the
demodulator.  
Let the inputs to the $\arctan\pqty{\cdot}$  block be the (real and
imaginary parts of) 
\begin{equation}
s_d\pqty{t} = A_d\pqty{t}\:e^{j\theta_d\pqty{t}}  
\end{equation}
where $A_d\pqty{t}$ is the possibly time-varying amplitude, and
$\theta_d\pqty{t}$ is the phase. The output $\theta_e\pqty{t}$ of the
$\arctan\pqty{\cdot}$ block is simply the phase $\theta_d\pqty{t}$.
Then, we have
\begin{equation}
\begin{aligned}
S_d\pqty{s} &= {\cal L}\Bqty{A_d\pqty{t}\:e^{j\theta_d\pqty{t}}} \\ 
&= A_o\,H_L\pqty{s}\,H_R^{BB}\pqty{s}\,S_{inr}^{BB}\pqty{s}\\
   &= \frac{A_o^2}{2} \,H_L\pqty{s}\,H_R^{BB}\pqty{s}\,{\cal
  L}\Bqty{e^{j\theta_o\pqty{t}}}
\end{aligned}
\label{eqn:demodfinalout}  
\end{equation}
Thus, we have obtained a compact and simple model for the entire
signal chain from the phase deviation $\theta_o\pqty{t}$ of the {\sc
  NCO} to the phase error $\theta_e\pqty{t}$, output of the phase
detector. In doing so, we were able to capture everything with
low-frequency signals, in a base-band equivalent manner. That is, the
model in \eqref{eqn:demodfinalout} does not have any explicit
frequency translation operations, making it {\it
  time-invariant}. However, this model is still {\it nonlinear} due to
the phase-to-complex conversion, i.e., $e^{j\cdot}$, in the {\sc NCO},
and the complex-to-phase conversion, i.e., $\arctan\pqty{\cdot}$, in
the demodulator. Next, we introduce a further approximation in order
to obtain a simple, {\it linear} and {\it time-invariant} model for
the entire signal chain from $\theta_o\pqty{t}$ to $\theta_e\pqty{t}$.

We first observe that the $\arctan\pqty{\cdot}$ block makes any
scaling or DC gain factor up to that point along the signal chain
irrelevant, i.e., the factor $A_o/2$ in
\eqref{eqn:resonatorinputlaplaceBB}, the DC gain
$Q/\pqty{m\omega_o^2}$ in \eqref{eqn:resonatorfreqrespBB5}, any DC
gain in $H_L\pqty{s}$, and the factor $A_o$ in \eqref{eqn:demodoutput}
are all immaterial for the final output of the demodulator. The final
operation in the demodulator subtracts the phase set point from the
computed phase, and is set to $-\pi/2$ due to the phase shift in
\eqref{eqn:resonatorfreqrespBB5} due to the resonator. (Please note
that the $-\pi/2$ phase set point is, not related to, and distinct
from the $\pi/2$ phase shift applied to the {\sc NCO} signal for the
quadrature arm of the demodulator.) Here, we assume that $H_L\pqty{s}$
does not introduce any extra phase shift for (complex-valued) DC
signals.  In this case, if the phase deviation $\theta_o\pqty{t}$ of
the {\sc NCO} is time-invariant, set to a constant as
$\theta_o\pqty{t}=\theta_c$, then the final output of the demodulator,
i.e., the phase error $\theta_e\pqty{t}$, is simply equal to this
constant phase $\theta_c$. Thus, we remove all scaling factors, DC
gains, as well as the $-\pi/2$ phase shift in resonator, from the
signal path, without changing the final phase error output of the
demodulator. We define
\begin{equation}
H_{R}\pqty{s} 
=\frac{1}{1+s\,\tau_r}\\
\label{eqn:resonatortranfunBBsimple}
\end{equation}
which was obtained from \eqref{eqn:resonatorfreqrespBB5} by removing
the DC gain and the $-\pi/2$ phase shift. We assume that the low-pass
filter $H_L\pqty{s}$ has a DC gain of 1 and introduces no phase shift
for (complex-valued) DC signals, i.e., $H_{L}\pqty{0} = 1$.
We then modify \eqref{eqn:demodfinalout} to obtain 
\begin{equation}
\begin{aligned}
{\cal L}\Bqty{A_d\pqty{t}\:e^{j\theta_d\pqty{t}}} = H_L\pqty{s}\,H_R\pqty{s}\,{\cal
  L}\Bqty{e^{j\theta_o\pqty{t}}}
\end{aligned}
\label{eqn:demodfinaloutmod}  
\end{equation}
We next assume that the phase deviation $\theta_o\pqty{t}$ is 
small enough so that we can use the following
approximation  
\begin{equation}
e^{j\theta_o\pqty{t}} \approx 1 + j\theta_o\pqty{t}
\label{eqn:phlinearization}  
\end{equation}
The constant DC factor 1 above (real part) will go through
$H_R\pqty{s}$ and $H_L\pqty{s}$ unmodified since $H_{L}\pqty{0} =
H_{R}\pqty{0} = 1$, whereas $j\theta_o\pqty{t}$ will be modified by
the dynamics of these filters, producing
\begin{equation}
1 + j\theta_d\pqty{t} \approx e^{j\theta_d\pqty{t}} 
\label{eqn:phlinearizationout}  
\end{equation}
where 
\begin{equation}
1 + j\,{\cal L}\Bqty{\theta_d\pqty{t}} = 1 + j\, H_L\pqty{s}\,H_R\pqty{s}\,{\cal
  L}\Bqty{\theta_o\pqty{t}}  
\label{eqn:phtranfun}  
\end{equation}
Thus, the phase error $\theta_e\pqty{t}=\theta_d\pqty{t}$ (ignoring
$-\pi/2$ phase set point) can be computed with
\begin{equation}
\begin{aligned}
\Theta_e\pqty{s} = {\cal L}\Bqty{\theta_e\pqty{t}} &= H_L\pqty{s}\,H_R\pqty{s}\,{\cal
  L}\Bqty{\theta_o\pqty{t}}\\ &= H_L\pqty{s}\,H_R\pqty{s}\,\Theta_o\pqty{s}  
\end{aligned}
\label{eqn:pherrtranfun}  
\end{equation}
Since $\Theta_o\pqty{s} = \tfrac{1}{s}\,\Delta\Omega\pqty{s}$, we have 
\begin{equation}
\Theta_e\pqty{s} = H_L\pqty{s}\,H_R\pqty{s}\,\frac{1}{s}\,\Delta\Omega\pqty{s}  
\label{eqn:freqdevtopherrtranfun}  
\end{equation}
Thus, we have obtained a simple,
{\it linear} and {\it time-invariant} model for the entire signal
chain from the frequency deviation $\Delta\Omega\pqty{t}$ to the phase
error $\theta_e\pqty{t}$. 

\begin{figure*}[t]
\centering
\includegraphics[width=0.99\textwidth]{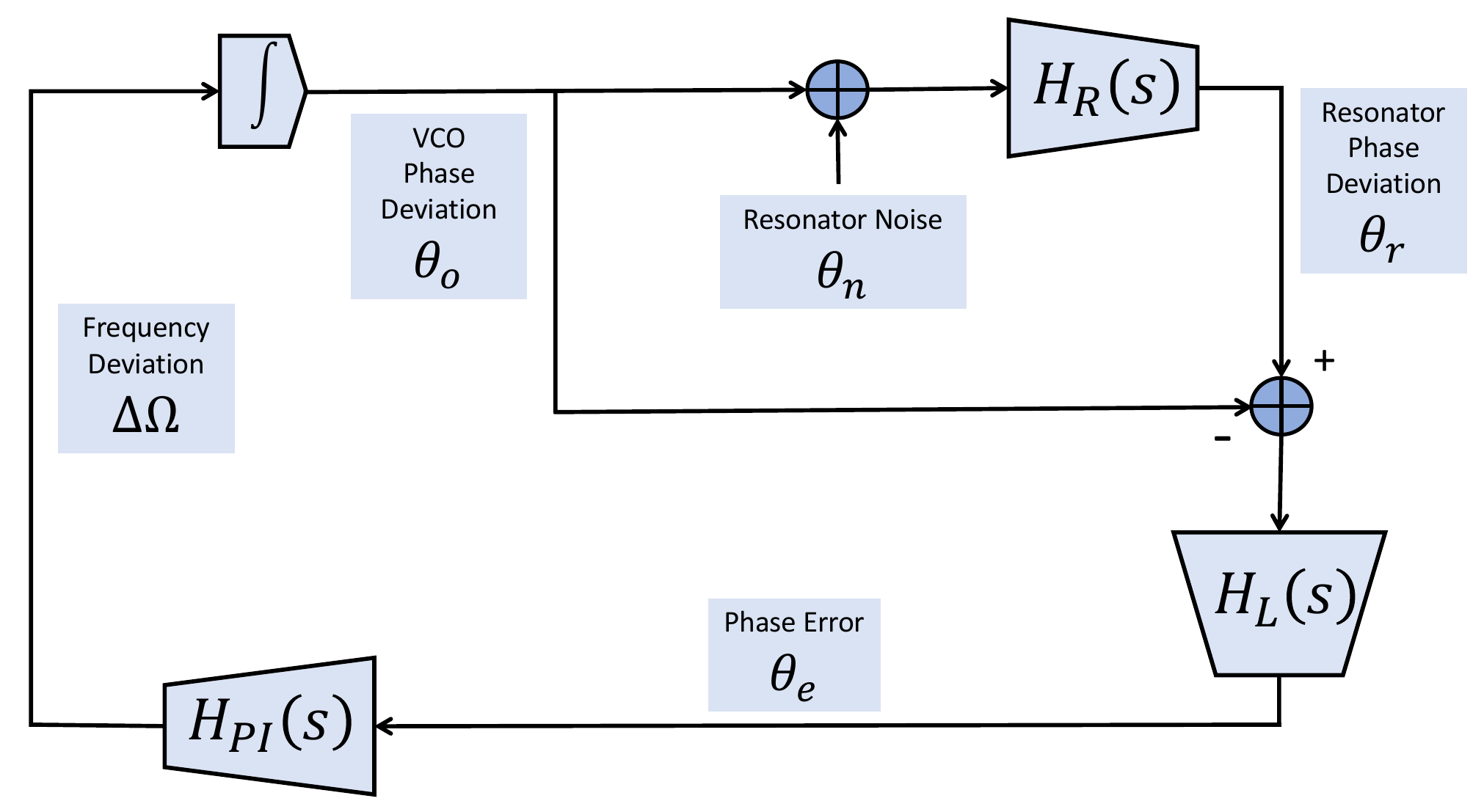}
\caption{Phase-domain base-band equivalent model of PLL based resonator tracking}
\label{fig:pllphase}
\end{figure*}

In deriving the model above, we assumed that the second demodulator
input is simply set to $A_o\cos\pqty{\omega_0t}$, as in the open-loop
model in Figure~\ref{fig:pllopen}. In the closed-loop {\sc PLL}, this
input is in fact set to the output of the {\sc NCO}, making the
demodulator effectively a {\it phase difference detector} between the
resonator and the {\sc NCO} outputs. We now construct a base-band
equivalent, phase domain model for the closed-loop {\sc PLL}, as shown
in Figure~\ref{fig:pllphase}, by taking this into account. In this
model, the PI controller is represented as a transfer function
$H_{PI}\pqty{s}$ given in \eqref{eqn:picontrollertransfun}. We have
verified all of the approximations we have performed in deriving the
phase domain model against simulations of the full nonlinear,
time-varying system model. However, we emphasize that, for the model
in Figure~\ref{fig:pllphase} to be valid, the base-band equivalent
resonator transfer function $H_R\pqty{s}$ and the demodulator low-pass
filter transfer function $H_L\pqty{s}$ need to satisfy $H_{L}\pqty{0}
= H_{R}\pqty{0} = 1$.

\subsection{Baseband equivalent noise model for the resonator}
\label{sec:noisemodel}
Having derived a simple model for the deterministic dynamics of the
system, we now turn to doing the same for the noise dynamics.  We
consider the open-loop system in Figure~\ref{fig:pllopen} and
initially set the {\sc NCO} output to zero. Thus, the
resonator-demodulator chain is driven by only the resonator noise
source, modeled as a stationary, white Gaussian random process with a
(two-sided) {\sc PSD} as given in \eqref{eqn:thmnoisePSD}. This white
noise source is shaped by the resonator, turning into a colored noise
process with a pass-band {\sc PSD}, but still stationary. In the
demodulator, it goes through the two mixers (multipliers), turning
into {\it cyclo-stationary} noise
processes~\cite{GardnerRP,RoLoFe97jssc,demir1998Kluwer}. As we will see later,
the low-pass filters in the demodulator not only block the
high-frequency parts but also {\it stationarize} these
cyclo-stationary noise processes by removing the high-order
cyclo-stationary
components~\cite{GardnerRP,RoLoFe97jssc,demir1998Kluwer}. Finally, the noise
processes in the in-phase (real) and quadrature (imaginary) arms of
the demodulator converge at the $\arctan\pqty{\cdot}$ nonlinearity,
which performs a real/imaginary-to-phase conversion, yielding a phase
error noise process at the very end. The phase set point subtraction
is a DC operation, and was taken into account as part of the
deterministic dynamics of the system considered before. As summarized
here, the thermo-mechanical noise of the resonator goes through a
nonlinear and time-varying system with inherent frequency translation
operations. However, as we did for the deterministic dynamics, we will
be able to model the entire noisy dynamics as captured by a much
simpler system, where an equivalent {\it stationary} noise process
passes through base-band equivalent {\it linear} and {\it
  time-invariant} filters.

We follow the {\it noise path} in Figure~\ref{fig:pllopen}
starting from the input of the resonator. With the {\sc PSD} of the
noise source at the resonator input as in \eqref{eqn:thmnoisePSD}, the
noise {\sc PSD} at the resonator output can be computed as follows
\begin{equation}
\begin{aligned}
S_{r}\pqty{\omega} & =
\vqty{H_{R}^{BP}\pqty{j\,\omega}}^2\;S_{thm}\pqty{\omega}\\
& = \frac{2\Gamma k_{B}\,T}{m}\frac{1}{\pqty{\omega^2-\omega_o^2}^2+\Gamma^2\,\omega^2}
\end{aligned}
\label{eqn:resoutnoisePSD}  
\end{equation}
Next, this pass-band stationary noise process is fed into the two
multipliers that generate cyclo-stationary noise, which can not be
characterized with a simple {\sc PSD}.  For cyclo-stationary
processes, the {\sc PSD} is a function of two variables, the frequency
$\omega$ and time $t$, i.e., $S_{cyc}\pqty{t,\omega}$, where the $t$
dependence is periodic and can be represented with a Fourier series as
discussed in
Apppendix~\ref{sec:specPSDcyclo}~\cite{GardnerRP,RoLoFe97jssc,demir1998Kluwer}.
The noise signal at the output of the in-phase (real part) multiplier
is given by
\begin{equation}
s_{mRe}\pqty{t} = A_o\cos\pqty{\omega_o t}\:s_{r}\pqty{t}  
\label{eqn:cycnosinphase}  
\end{equation}
where $s_{r}\pqty{t}$ is the stationary noise signal at the output of the
resonator with the {\sc PSD} in \eqref{eqn:resoutnoisePSD}.
Then, the {\it cyclic spectra} of $s_{mRe}\pqty{t}$, as shown in
Apppendix~\ref{sec:specPSDcyclo}, is given by 
\begin{equation}
\begin{aligned}
S_{mRe}^{\pqty{0}}\pqty{\omega} &=
\frac{A_o^2}{4}\bqty{S_{r}\pqty{\omega-\omega_o}+S_{r}\pqty{\omega+\omega_o}}\\
 S_{mRe}^{\pqty{2}}\pqty{\omega} & =  S_{mRe}^{\pqty{-2}}\pqty{\omega} =
\frac{A_o^2}{4}S_{r}\pqty{\omega}\\
 S_{mRe}^{\pqty{k}}\pqty{\omega} &= 0\quad \text{for all other}\;\;k
\end{aligned}
\label{eqn:cycnosinphasePSD}  
\end{equation}
The noise signal at the output of the quadrature (imaginary part) multiplier is given by 
\begin{equation}
s_{mIm}\pqty{t} = -A_o\sin\pqty{\omega_o t}\:s_{r}\pqty{t}  
\label{eqn:cycnosquad}  
\end{equation}
It can be easily shown that the {\it cyclic spectra} of $s_{mIm}\pqty{t}$ is  
\begin{equation}
\begin{aligned}
S_{mIm}^{\pqty{0}}\pqty{\omega} &=
\frac{A_o^2}{4}\bqty{S_{r}\pqty{\omega-\omega_o}+S_{r}\pqty{\omega+\omega_o}}\\
 S_{mIm}^{\pqty{2}}\pqty{\omega} & =  S_{mIm}^{\pqty{-2}}\pqty{\omega} =
-\frac{A_o^2}{4}S_{r}\pqty{\omega}\\
 S_{mIm}^{\pqty{k}}\pqty{\omega} &= 0\quad \text{for all other}\;\;k
\end{aligned}
\label{eqn:cycnosquadPSD}  
\end{equation}
The cyclo-stationary noise signals $s_{mRe}\pqty{t}$ and
$s_{mIm}\pqty{t}$ at the outputs of the multipliers are filtered with
the low-pass filter $H_L\pqty{s}$. As shown in
Apppendix~\ref{sec:specPSDcyclo}, this filter {\it stationarizes}
these cyclo-stationary noise processes, and at the same time removes
high-frequency components~\cite{RoLoFe97jssc}. The
stationary noise signals, $s_{dRe}\pqty{t}$ and $s_{dIm}\pqty{t}$, at
the output of these filters have the following {\sc PSD}
\begin{equation}
\begin{aligned}
S_{dRe}\pqty{\omega} &= S_{dIm}\pqty{\omega} =\\ 
&\frac{A_o^2}{4}\,\vqty{H_L\pqty{j\omega}}^2\,\bqty{S_{r}\pqty{\omega-\omega_o}+S_{r}\pqty{\omega+\omega_o}}
\end{aligned}
\label{eqn:psdemodfilterout}  
\end{equation}
where $S_{r}\pqty{\omega}$ is the {\sc PSD} in
\eqref{eqn:resoutnoisePSD}. We next analyze the noise folding
(discussed in Apppendix~\ref{sec:specPSDcyclo}) and filtering
represented by \eqref{eqn:psdemodfilterout}.  $S_{r}\pqty{\omega}$ is
a pass-band, two-sided {\sc PSD} with power concentrated around
$\pm\,\omega_o$. Thus, $S_{r}\pqty{\omega-\omega_o}$ has power
concentrated around 0 and $2\omega_o$, whereas for
$S_{r}\pqty{\omega+\omega_o}$ it is around 0 and
$-2\omega_o$. Assuming that $H_L\pqty{j\omega}$ is a low-pass filter
with an effective bandwidth that is much less than $\omega_o$,
satisfying $H_L\pqty{\pm j2\omega_o} \approx 0$, it will remove the
noise component at $2\omega_o$ in $S_{r}\pqty{\omega-\omega_o}$ and
the one at $-2\omega_o$ in $S_{r}\pqty{\omega+\omega_o}$. Then, the
only noise components of interest are the ones around 0. 
We evaluate these components as follows. 
We first rewrite \eqref{eqn:resoutnoisePSD} as below 
\begin{equation}
S_{r}\pqty{\omega} 
 = \frac{2\Gamma k_{B}\,T}{m}\frac{1}{\bqty{\pqty{\omega+\omega_o}\pqty{\omega-\omega_o}}^2+\Gamma^2\,\omega^2}
\label{eqn:resoutnoisePSDmodi}  
\end{equation}
and then 
\begin{equation}
\begin{aligned}
 S_{r}\pqty{\omega-\omega_o}&+S_{r}\pqty{\omega+\omega_o} =  \\
\frac{2\Gamma k_{B}\,T}{m}&\left[
  \frac{1}{\bqty{\omega\pqty{\omega-2\omega_o}}^2+\Gamma^2\,\pqty{\omega-\omega_o}^2}\right.\\
&\left.+\frac{1}{\bqty{\omega\pqty{\omega+2\omega_o}}^2+\Gamma^2\,\pqty{\omega+\omega_o}^2}\right]
\end{aligned}
\label{eqn:psdemodoutsimpl}
\end{equation}
We then assume that $\omega\ll\omega_o$ and use the following
approximations due to the fact that we are interested in 
the above {\sc PSD} only at low frequencies  
\begin{equation}
\omega\pm 2\,\omega_o \approx \pm 2\,\omega_o\quad,\quad
\omega\pm \omega_o \approx \pm \omega_o 
\label{eqn:approxforPSD}
\end{equation}
The above approximations are similar to the ones in
\eqref{eqn:approxforfregresp} that we employed in simplifying the
deterministic dynamics. 
With \eqref{eqn:approxforPSD}, \eqref{eqn:psdemodoutsimpl} can be
simplified as follows 
\begin{equation}
\begin{aligned}
S_{r}\pqty{\omega-\omega_o}\!\!\!&\:\:+S_{r}\pqty{\omega+\omega_o} \approx  \\
\frac{2\Gamma k_{B}\,T}{m}&\bqty{
  \frac{1}{\pqty{2\omega\omega_o}^2+\Gamma^2\,\omega_o^2}
+\frac{1}{\pqty{2\omega\omega_o}^2+\Gamma^2\,\omega_o^2}}=\\
\frac{4\Gamma k_{B}\,T}{m}&
  \frac{1}{\pqty{2\omega\omega_o}^2+\Gamma^2\,\omega_o^2}=\\
\frac{4 k_{B}\,T}{m\,\Gamma\omega_o^2}& 
\frac{1}{1 + \pqty{\frac{2\,\omega}{\Gamma}}^2} = 
\frac{4 k_{B}\,T}{m\,\Gamma\omega_o^2}\frac{1}{1 + \pqty{\omega\,\tau_r}^2} 
\end{aligned}
\label{eqn:psdemodoutsimpl2}
\end{equation}
We substitute \eqref{eqn:psdemodoutsimpl2}  above into
\eqref{eqn:psdemodfilterout} to obtain  
\begin{equation}
\begin{aligned}
S_{dRe}\pqty{\omega} &= S_{dIm}\pqty{\omega} =\\
&\frac{A_o^2}{4}\,\vqty{H_L\pqty{j\omega}}^2\, 
\frac{4 k_{B}\,T}{m\,\Gamma\omega_o^2}\frac{1}{1 + \pqty{\omega\,\tau_r}^2}
\end{aligned}
\label{eqn:psdemodfilteroutsimple1}  
\end{equation}
We observe that 
\begin{equation}
\frac{1}{1 + \pqty{\omega\,\tau_r}^2} = \vqty{H_{R}\pqty{j\omega}}^2 
\end{equation}
with $H_{R}\pqty{s}$ defined as in
\eqref{eqn:resonatortranfunBBsimple}. Thus, 
\begin{equation}
\begin{aligned}
S_{dRe}\pqty{\omega} = S_{dIm}\pqty{\omega} =
\frac{A_o^2\,k_{B}\,T}{m\,\Gamma\omega_o^2}\vqty{H_L\pqty{j\omega}}^2\vqty{H_{R}\pqty{j\omega}}^2 
\end{aligned}
\label{eqn:psdemodfilteroutsimple}  
\end{equation}
Final stage in the demodulator is the $\arctan\pqty{\cdot}$ block, a
memoryless nonlinearity. Up till now, we assumed that the
resonator-demodulator chain is driven by only the resonator noise
source. In order to correctly evaluate the effect of the nonlinear
$\arctan\pqty{\cdot}$ block, we need to also consider the signal input
to the resonator that is fed from the {\sc NCO} output. With {\sc NCO}
output set to
\begin{equation}
s_o\pqty{t}  = A_o\cos\pqty{\omega_o\,t }
 =  \frac{A_o}{2}\bqty{e^{j \omega_o t}+e^{-j \omega_o t}}
\label{eqn:NCOsignalout} 
\end{equation} 
the deterministic components of the in-phase and quadrature signals at
the inputs of the $\arctan\pqty{\cdot}$ block will be the real and
imaginary parts of 
\begin{equation}
\frac{A_o^2}{2}\frac{Q}{m\,\omega_o^2}e^{-j\frac{\pi}{2}} = 0 -j\frac{A_o^2}{2}\frac{Q}{m\,\omega_o^2} 
\label{eqn:determataninp}  
\end{equation}
based on \eqref{eqn:demodoutdet} and
\eqref{eqn:resonatorfreqrespBB5}. 
The output $\theta_d\pqty{t}$ of the $\arctan\pqty{\cdot}$ can be
computed as follows 
\begin{equation}
\theta_d\pqty{t} = \arctan \frac{-\frac{A_o^2}{2}\frac{Q}{m\,\omega_o^2}+s_{dIm}\pqty{t}}{s_{dRe}\pqty{t}}
\label{eqn:atanout}  
\end{equation}
where $s_{dRe}\pqty{t}$ and $s_{dIm}\pqty{t}$ are the noise signals at
the inputs of $\arctan\pqty{\cdot}$. We observe that the noise signal $s_{dIm}\pqty{t}$
above is much smaller than the DC signal term
$-\tfrac{A_o^2}{2}\tfrac{Q}{m\,\omega_o^2}$. Thus,
\begin{equation}
\theta_d\pqty{t} \approx \arctan \frac{-\frac{A_o^2}{2}\frac{Q}{m\,\omega_o^2}}{s_{dRe}\pqty{t}}
\label{eqn:atanoutsimp1}  
\end{equation}
Furthermore, the noise term $s_{dRe}\pqty{t}$ is also small. Thus, we
use the following first-order Taylor's series expansion 
\begin{equation}
\arctan \frac{a}{x} \approx -\frac{\pi}{2}-\frac{x}{a}\quad \text{for small}\;\;x
\label{eqn:fotaylorarctan}  
\end{equation}
Hence, we have 
\begin{equation}
\theta_d\pqty{t} \approx \arctan
\frac{-\frac{A_o^2}{2}\frac{Q}{m\,\omega_o^2}}{s_{dRe}\pqty{t}} \approx
-\frac{\pi}{2} + \frac{2\,m\,\omega_o^2}{A_o^2\,Q}\:s_{dRe}\pqty{t}
\label{eqn:atanoutsimp2}  
\end{equation}
With the subtraction of the phase set point, i.e., $-\pi/2$ at the
output of the demodulator, the phase error $\theta_e\pqty{t}$ is given
by 
\begin{equation}
\theta_e\pqty{t} = \theta_d\pqty{t} - \pqty{ -\frac{\pi}{2}} = \frac{2\,m\,\omega_o^2}{A_o^2\,Q}\:s_{dRe}\pqty{t}
\label{eqn:phaserrno}  
\end{equation}
Then, the {\sc PSD} of $\theta_e\pqty{t}$ can be computed based on
\eqref{eqn:psdemodfilteroutsimple} and \eqref{eqn:phaserrno}:
\begin{equation}
\begin{aligned}
S_{\theta_e}\pqty{\omega} & = \pqty{\frac{2\,m\,\omega_o^2}{A_o^2\,Q}}^2
\frac{A_o^2\,k_{B}\,T}{m\,\Gamma\omega_o^2}\,\vqty{H_L\pqty{j\omega}}^2\,\vqty{H_{R}\pqty{j\omega}}^2
\\
& = \frac{4\,m\,\omega_o^2\,k_B\, T}{A_o^2\,Q^2\,\Gamma} \,\vqty{H_L\pqty{j\omega}}^2\,\vqty{H_{R}\pqty{j\omega}}^2
\end{aligned}
\label{eqn:psdphaserr}  
\end{equation}
We use \eqref{eqn:resonatorTC} in \eqref{eqn:psdphaserr} and define 
\begin{equation}
\begin{aligned}
S_{\theta_n}\pqty{\omega} = \frac{4\,m\,\omega_o^2\,k_B\,
  T}{A_o^2\,Q^2\,\Gamma} = \frac{2\,m\,\tau_r\,\omega_o^2\,k_B\,  T}{A_o^2\,Q^2} 
\end{aligned}
\label{eqn:psdthetaresnos}  
\end{equation}
as the {\sc PSD} of a white, Gaussian noise process $\theta_n\pqty{t}$.
$\theta_n\pqty{t}$
represents the thermo-mechanical noise of the resonator, in
an input-referred manner, in the base-band equivalent phase domain
model in Figure~\ref{fig:pllphase}. This noise process goes through
two filters, $H_R\pqty{s}$ and $H_R\pqty{s}$, as in both
\eqref{eqn:psdphaserr} and Figure~\ref{fig:pllphase}, to produce the
phase error noise at the output of the demodulator, i.e., the input of
the {\sc PI} controller

\subsection{{\sc \it PLL} noise analysis}

Having derived base-band equivalent phase domain models for both the
deterministic and noise dynamics of the {\sc PLL} components, we next
proceed with the noise analysis of the {\it closed-loop} system based
on Figure~\ref{fig:pllphase}. 

The closed-system is governed by the following equation, written
directly in the frequency domain, by going around the loop in
Figure~\ref{fig:pllphase}:
\begin{equation}
\Theta_o\pqty{s} = \frac{1}{s} H_{PI}\pqty{s}H_L\pqty{s}\bqty{H_R\pqty{s}\bqty{\Theta_n\pqty{s}+\Theta_o\pqty{s}}-\Theta_o\pqty{s}} 
\label{eqn:plleqn}  
\nonumber
\end{equation}
We compute the transfer function from $\Theta_n\pqty{s}$ to
$\Theta_o\pqty{s}$ by solving the loop equation above, and substitute 
\eqref{eqn:picontrollertransfun} and
\eqref{eqn:resonatortranfunBBsimple} into the result to obtain 
\begin{equation}
\begin{aligned}
H_{\theta_n}^{\theta_o}\pqty{s} = \frac{\Theta_o\pqty{s}}{\Theta_n\pqty{s}} & =
\frac{H_{PI}\pqty{s}H_L\pqty{s}H_R\pqty{s}}{s+H_{PI}\pqty{s}H_L\pqty{s}\bqty{1-H_R\pqty{s}}}\\
& =
\frac{\pqty{K_p+\frac{K_i}{s}}H_L\pqty{s}\frac{1}{1+s\tau_r}}{s+\pqty{K_p+\frac{K_i}{s}}
  H_L\pqty{s} \frac{s\tau_r}{1+s\tau_r}}
\end{aligned}
\label{eqn:plltrfunfromthetntotheto_pre}  
\end{equation}
We manipulate the above expression to obtain
\begin{equation}
\begin{aligned}
H_{\theta_n}^{\theta_o}\pqty{s}=\frac{1}{s}\frac{1}{\tau_r}\bqty{
\frac{\pqty{s K_p+ K_i} H_L\pqty{s}}{s^2+\frac{s}{\tau_r}+\pqty{s K_p+ K_i} H_L\pqty{s}}}
\end{aligned}
\label{eqn:plltrfunfromthetntotheto}  
\end{equation}
Due to \eqref{eqn:freqtophase}, 
the transfer function from $\Theta_n\pqty{s}$ to the {\sc NCO}
frequency deviation $\Delta\Omega\pqty{s}$ is 
\begin{equation}
\begin{aligned}
H_{\theta_n}^{\Delta\Omega}\pqty{s}= \frac{1}{\tau_r}\bqty{
\frac{\pqty{s K_p+ K_i} H_L\pqty{s}}{s^2+\frac{s}{\tau_r}+\pqty{s K_p+ K_i} H_L\pqty{s}}}
\end{aligned}
\label{eqn:plltrfunfromthetntodeltaomega}  
\end{equation}
We can then compute the {\sc PSD} for the frequency deviation $\Delta\Omega\pqty{t}$ of
the {\sc NCO} as follows 
\begin{equation}
\begin{aligned}
S_{\Delta\Omega}\pqty{\omega} =\vqty{H_{\theta_n}^{\Delta\Omega}\pqty{j\omega}}^2\, S_{\theta_n}\pqty{\omega} 
\end{aligned}
\label{eqn:psdphaseandfreqdev}
\end{equation}
with $S_{\theta_n}\pqty{\omega}$ as given in
\eqref{eqn:psdthetaresnos}. 
We note that the transfer function in
\eqref{eqn:plltrfunfromthetntodeltaomega} satisfies 
\begin{equation}
H_{\theta_n}^{\Delta\Omega}\pqty{s\rightarrow 0} = \frac{1}{\tau_r}\quad,\quad H_{\theta_n}^{\Delta\Omega}\pqty{s\rightarrow\pm\infty} = 0  
\label{eqn:plltrfunfromthetntodeltaomegas0}  
\end{equation}
with an appropriate low-pass filter $H_L\pqty{s}$ in the
demodulator. That is, $H_{\theta_n}^{\Delta\Omega}\pqty{s}$ represents
a low-pass filter, with a bandwidth that is essentially the {\it loop
  bandwidth} for the {\sc PLL}. With $S_{\theta_n}\pqty{\omega}$ in
\eqref{eqn:psdthetaresnos} representing a white spectrum and due to
\eqref{eqn:psdphaseandfreqdev}, the frequency deviation
$\Delta\Omega\pqty{t}$ has the characteristics of band-limited
(low-pass filtered) white noise. The phase deviation
$\theta_o\pqty{t}$, the (time) integral of $\Delta\Omega\pqty{t}$,
then has the characteristics of a {\it random walk}, albeit not in the
form of a standard Wiener process (Brownian motion). However, for
longer time scales (larger than the {\sc PLL} loop time constant)
$\theta_o\pqty{t}$ does behave as a standard random walk process. We
emphasize here that this random walk aspect of the phase deviation is
not arising from the inherent phase noise of the {\sc VCO} or {\sc
  NCO}. We recall that we have set the inherent phase noise of the
controlled oscillator to zero, upfront, when we started our analysis.
The random walk nature of phase deviation we have derived above is due
to the fact the thermo-mechanical noise of the resonator {\it
  circulates around the loop}. One can interpret that the loop dynamics {\it
  converts} the additive, thermo-mechanical noise (in amplitude) of
the resonator into phase noise in the {\sc NCO}. Our analysis above
reveals, in a rigorous manner, precisely how this conversion occurs.
This resultant phase noise in the {\sc NCO} ultimately limits the
frequency tracking accuracy of the {\sc PLL} and represents a
fundamental limit on the sensitivity of the resonator based sensor
system. Based on the model and theory we have developed above, we next
precisely characterize the frequency tracking accuracy of the {\sc
  PLL} architecture in terms of Allan Deviation.

\subsection{Characterizing {\sc \it PLL} performance via Allan
  Deviation} {\it Allan
  Deviation}~\cite{allan1966statistics,rubiola2009phase,wiki:Allanvariance}
is the standard measure of frequency stability.  It is widely used in
assessing the sensitivity of resonant sensors.  Please
see~\cite{allan1966statistics,rubiola2009phase,wiki:Allanvariance} for
details on Allan Deviation. Our discussion is based
on~\cite{allan1966statistics,rubiola2009phase,wiki:Allanvariance}.
 
We define $y\pqty{t}$
to be the {\it fractional frequency deviation} as follows
\begin{equation}
 y\pqty{t} = \frac{\Delta\Omega\pqty{t}}{\omega_o}  
\label{eqn:fracfreqdev}  
\end{equation}
where $\Delta\Omega\pqty{t}$ is the {\sc NCO} frequency deviation, and
$\omega_o$ is the nominal {\sc NCO} frequency. The {\it averaged
  fractional frequency deviation} $\bar{y}\pqty{t,\tau}$ is defined as 
\begin{equation}
\bar{y}\pqty{t,\tau} = \frac{1}{\tau} \int_0^\tau y\pqty{t+u}\dd u 
\label{eqn:avfracfreqdev}  
\end{equation}
where $\tau$ is the {\it averaging time}. 
The $i$th sample of $\bar{y}\pqty{t,\tau}$ is given by 
\begin{equation}
\bar{y}_i = \bar{y}\pqty{i\tau,\tau}
\label{eqn:sampledavfracfreqdev}  
\end{equation}
where the sampling interval is chosen to be equal to the averaging
time $\tau$ for standard Allan Deviation.  Finally, {\it Allan Variance} is
computed as follows
\begin{equation}
\sigma_y^2\pqty{\tau} = \frac{1}{2}\:\vb{E}\bqty{\pqty{\bar{y}_{i+1}-\bar{y}_i}^2} 
\label{eqn:allanvar}  
\end{equation}
where where $ \vb{E}\bqty{\cdot}$ denotes the probabilistic
expectation operator. The {\it Allan Deviation} is then 
\begin{equation}
\sigma_y\pqty{\tau} =  \sqrt{\sigma_y^2\pqty{\tau}}
\label{eqn:allandev}  
\end{equation}
We note that the above definition implicitly assumes or postulates
that $\sigma_y^2\pqty{\tau}$ is a function of only the averaging time
$\tau$, and is independent of the sampling times represented by $i$
and $i+1$. This is the case when $y\pqty{t}$ is a (wide-sense)
stationary process. This is satisfied in our setting, since the
frequency deviation $\Delta\Omega\pqty{t}$ is indeed a stationary
process, as the output of a stable, linear and time-invariant system
(with transfer function $H_{\theta_n}^{\Delta\Omega}\pqty{s}$ in
\eqref{eqn:plltrfunfromthetntodeltaomega}) with its input set to
white, stationary, Gaussian noise $\theta_n\pqty{t}$.

It can be shown that (as in Appendix~\ref{sec:allendev})
\begin{equation}
\sigma_y^2\pqty{\tau} = \frac{4}{\pi\tau^2}\:\int_{-\infty}^{+\infty}
\frac{\bqty{\sin\pqty{\frac{\omega\,\tau}{2}}}^4}{\omega^2}\:S_y\pqty{\omega}\,\dd \omega  
\label{eqn:allanvarfrompsd}  
\end{equation}
where $S_y\pqty{\omega}$ is the {\sc PSD} of $y\pqty{t}$ given by 
\begin{equation}
S_y\pqty{\omega} = \frac{1}{\omega_o^2}S_{\Delta\Omega}\pqty{\omega} 
\label{eqn:psdy}  
\end{equation}
due to \eqref{eqn:fracfreqdev} with $S_{\Delta\Omega}\pqty{\omega}$ in
\eqref{eqn:psdphaseandfreqdev}.  With the transfer function in
\eqref{eqn:plltrfunfromthetntodeltaomega}, it is, unfortunately, not
possible to evaluate the Allan Variance integral in
\eqref{eqn:allanvarfrompsd} analytically. However, we can evaluate it
numerically in order to compute the Allan Deviation for all values of
$\tau$, for any choice of the low-pass filter $H_L\pqty{s}$ and the
controller parameters $K_p$ and $K_i$. We will present results for
this numerical evaluation in Section~\ref{sec:simulation}. On the
other hand, it is really desirable that we have an analytical handle
on the frequency tracking accuracy of the {\sc PLL} system. We next
evaluate the integral in \eqref{eqn:allanvarfrompsd} analytically, for
values of $\tau$ that are larger than the loop time constant, thus
computing a high-$\tau$ asymptote for the Allan Deviation of the {\sc
  PLL} tracking system. This result will be practically valuable,
since the {\sc PLL} is able to track the frequency deviations of the
resonator within its bandwidth, or in other words, at time scales that
are longer than the loop time constant. Allan Deviation for
values of $\tau$ larger than the loop time constant is important from
a practical point of view. Frequency deviations that occur faster than
the loop time constant are attenuated by the loop dynamics, rendering
the {\sc PLL} tracking system not useful at short time scales.

The high-$\tau$ asymptote for $\sigma_y^2\pqty{\tau}$ in
\eqref{eqn:allanvarfrompsd} can be computed by approximating
the low-pass {\sc PSD} $S_y\pqty{\omega}$ with its value at zero (low)
frequency, i.e., with 
\begin{equation}
S_y\pqty{\omega} \approx S_y\pqty{0} =
\frac{1}{\omega_o^2}S_{\Delta\Omega}\pqty{0} =
\frac{S_{\theta_n}\pqty{\omega}}{\omega_o^2\,\tau_r^2} = \frac{S_{\theta_n}}{\omega_o^2\,\tau_r^2}   
\label{eqn:ypsdlowfreq}  
\end{equation}
where we used \eqref{eqn:psdphaseandfreqdev} and
\eqref{eqn:plltrfunfromthetntodeltaomegas0}, with
$S_{\theta_n}\pqty{\omega} = S_{\theta_n}$ as a constant function of $\omega$ as
given in \eqref{eqn:psdthetaresnos}. Next, we substitute
\eqref{eqn:ypsdlowfreq}  in \eqref{eqn:allanvarfrompsd} and evaluate
the integral to obtain 
\begin{equation}
\begin{aligned}
\sigma_y^2\pqty{\tau} & = \frac{4}{\pi\tau^2}\:\int_{-\infty}^{+\infty}
\frac{\bqty{\sin\pqty{\frac{\omega\,\tau}{2}}}^4}{\omega^2}\:
\frac{S_{\theta_n}}{\omega_o^2\,\tau_r^2}\,\dd \omega \\
& = \frac{S_{\theta_n}}{\omega_o^2\,\tau_r^2}\,\frac{4}{\pi\tau^2}\:\int_{-\infty}^{+\infty}
\frac{\bqty{\sin\pqty{\frac{\omega\,\tau}{2}}}^4}{\omega^2}\,\dd \omega \\
& =
\frac{S_{\theta_n}}{\omega_o^2\,\tau_r^2}\,\frac{4}{\pi\tau^2}\,\frac{\pi\tau}{4}\\
& =
\frac{S_{\theta_n}}{\omega_o^2\,\tau_r^2}\,\frac{1}{\tau}
\end{aligned}
\label{eqn:allanvarlargetau}  
\end{equation}
We note that $\tau_r$ above is the resonator time constant defined by
\eqref{eqn:resonatorTC}, whereas $\tau$ is the averaging time used in
the definition of Allan Variance.  The above result for
$\sigma_y^2\pqty{\tau}$ is valid for
large $\tau$, larger than the loop time constant. We use 
\eqref{eqn:psdthetaresnos} and \eqref{eqn:resonatorTC} in 
\eqref{eqn:allanvarlargetau}:
\begin{equation}
\begin{aligned}
\sigma_y^2\pqty{\tau} &=
\frac{S_{\theta_n}}{\omega_o^2\,\tau_r^2}\,\frac{1}{\tau}\quad\text{for
  large}\:\tau\\
& = \frac{2\,m\,\tau_r\,\omega_o^2\,k_B\,
  T}{A_o^2\,Q^2}\,\frac{1}{\omega_o^2\,\tau_r^2}\,\frac{1}{\tau}\\
& = \frac{2\,m\,k_B\,
  T}{A_o^2\,Q^2\,\tau_r}\,\frac{1}{\tau}\\
& = \frac{m\,\omega_o\,k_B\,
  T}{A_o^2\,Q^3}\,\frac{1}{\tau}
\end{aligned}
\label{eqn:allanvarlargetausimple}  
\end{equation}
The factor (that multiplies $1/\tau$) above is expressed in terms of
the resonator parameters $m$, $\omega_o$, $Q$, Boltzmann's constant
$k_B$, temperature $T$, and the amplitude $A_o$ of the signal that
drives the resonator. 
We would like to express this factor 
in terms of a {\it Signal-to-Noise-Ratio} ($\text{\small\textsf{SNR}}$) for the
resonator. We define $\text{\small\textsf{SNR}}$ as follows
\begin{equation}
\begin{aligned}
  \text{\small \textsf{SNR}} & = \sqrt{\frac{\text{\small signal power at resonator
        input}}{\text{\small noise power
      at resonator input}}}\\
   & = \sqrt{\frac{A_o^2/2}{S_{thm}\,\text{\small\textsf{BW}}\,2}}\\
   & = \sqrt{\frac{A_o^2/2}{ 2\,m\,\Gamma\,k_{B}\,T\,\text{\small\textsf{BW}}\,2}}\\
   & = \sqrt{\frac{A_o^2\,Q}{ 8\,m\,\omega_o\,k_{B}\,T\,\text{\small\textsf{BW}}}}
\end{aligned}
\label{eqn:snrdef}  
\end{equation}
In the above, $S_{thm}$ is the input-referred, white (two-sided) {\sc
  PSD} of the thermo-mechanical noise of the resonator given in
\eqref{eqn:thmnoisePSD}. $\text{\small\textsf{BW}}$ is defined as the
(one-sided, hence the factor of 2) noise bandwidth.
$\text{\small\textsf{BW}}$ is typically set to the bandwidth of the
low-pass filters in the demodulator. Alternatively, it could be set to
the {\sc PLL} loop bandwidth. The particular choice for
$\text{\small\textsf{BW}}$ simply affects the
$\text{\small\textsf{SNR}}$ definition, there is nothing fundamental
about it.  
We note the following relationship for the product of $Q$ and
$\text{\small\textsf{SNR}}$:
\begin{equation}
Q\:\text{\small \textsf{SNR}}  =  \sqrt{\frac{A_o^2\,Q^3}{ 8\,m\,\omega_o\,k_{B}\,T\,\text{\small\textsf{BW}}}}
\label{eqn:snrQ}  
\end{equation}
Allan Variance in \eqref{eqn:allanvarlargetausimple} can be expressed
in terms of the product $Q\:\text{\small \textsf{SNR}}$: 
\begin{equation}
\begin{aligned}
\sigma_y^2\pqty{\tau} &= \frac{m\,\omega_o\,k_B\,
  T}{A_o^2\,Q^3}\,\frac{1}{\tau}\\
&= \frac{1}{8\,\pqty{Q\,\text{\small \textsf{SNR}}}^2\,\text{\small\textsf{BW}}}\,\frac{1}{\tau}
\end{aligned}
\label{eqn:allanvarlargetauQ}  
\end{equation} 
and Allan Deviation is 
\begin{equation}
\begin{aligned}
\sigma_y\pqty{\tau} = \frac{1}{2\sqrt{2}\,Q\,\text{\small
    \textsf{SNR}}\,\sqrt{\text{\small\textsf{BW}}}}\,\frac{1}{\sqrt{\tau}}
\end{aligned}
\label{eqn:allanvdevlargetauQ}  
\end{equation} 
The high-$\tau$ approximations above for Allan Variance and Deviation
with $1/\tau$ and $1/\sqrt{\tau}$ dependence represent random walk
phase noise~\cite{rubiola2009phase,wiki:Allanvariance}.  Indeed, for
time scales larger than the loop time constant, the resulting phase
deviation, arising from the thermo-mechanical noise of the resonator
and the loop dynamics, has a random walk nature. As we will see in
Section~\ref{sec:simulation}, Allan Deviation will have a different
dependence on $\tau$ for shorter time scales, however, still with the
same front (scaling) factor that has a $1/\pqty{Q\,\text{\small
    \textsf{SNR}}}$ form.  We note that the result in
\eqref{eqn:allanvdevlargetauQ}, valid for high-$\tau$, is independent
of the loop and controller parameters (apart from an indirect
dependence on them through $\text{\small\textsf{BW}}$ definition) such
as $K_p$ and $K_i$ and the particular choice for the filter transfer
function $H_L\pqty{s}$. On the other hand, these parameters do
determine the loop bandwidth and time constant, and how Allan
Deviation changes with $\tau$ for short time scales within the loop
time constant. However, we emphasize, the scaling factor
$1/\pqty{Q\,\text{\small \textsf{SNR}}}$ applies in this case as well.

%% file: simulation.tex
\section{Results versus Simulations}
\label{sec:simulation}
In developing the theory in Section~\ref{sec:theory}, we used several
approximations and assumptions:
\begin{list}{}{\leftmargin=0em \labelwidth=1em \itemindent=1em}
\item[$\bullet$] We assumed that the resonator output signal is a strictly
  band-limited band-pass signal, and that the low-pass filters in the
  demodulator/phase detector completely remove the high frequency
  signal and noise components produced by the multipliers. This
  allowed us to develop simpler, base-band equivalent models for the
  resonator and the phase detector.
\item[$\bullet$]  The base-band equivalent transfer function for the resonator and
  the thermo-mechanical noise {\sc PSD} were approximated as in
  \eqref{eqn:resonatorfreqrespBB3} and
  \eqref{eqn:psdemodoutsimpl2}. This allowed us to model the resonator
  with a one-pole low-pass transfer function in a base-band equivalent
  manner.
\item[$\bullet$]  We have used a linear(ized) model for the $\arctan\pqty{\cdot}$
  nonlinearity for both the deterministic and noisy dynamics of the
  {\sc PLL}, as in \eqref{eqn:phlinearization},
  \eqref{eqn:phlinearizationout} and \eqref{eqn:fotaylorarctan}. This
  allowed us to derive a {\it phase domain}, in addition to base-band
  equivalent, model for the loop dynamics.
\item[$\bullet$]  In deriving the deterministic and noise models for the resonator
  and the phase detector, we used the open-loop setting in
  Figure~\ref{fig:pllopen}, where the second demodulator input was set
  to $A_o\cos\pqty{\omega_0t}$. In the closed-loop system, this
  input comes from the {\sc NCO} and is equal to
  $A_o\cos\pqty{\omega_o\,t+\theta_o\pqty{t}}$, including the phase
  deviation $\theta_o\pqty{t}$. In the closed-loop phase domain  model in   
  Figure~\ref{fig:pllopen}, we took this into account by feeding the
  second demodulator input from the {\sc NCO} output. However, the
  base-band equivalent resonator noise model was derived in
  Section~\ref{sec:noisemodel} based on the assumption that the second
  demodulator input does not have any phase deviation. This simplified
  the noise model derivation considerably.   
\end{list}
Even though the above assumptions and approximations are well founded
and justified, we still would like to verify them. We do this by
comparing our analytical results against the ones obtained from
extensive, carefully crafted and run, time-domain stochastic
simulations of the {\sc PLL} system. In these simulations, none of the
above assumptions and approximations are used. The system is simulated
with full, high-frequency, nonlinear and time-varying models for the
resonator and the demodulator as shown in Figure~\ref{fig:pllfull}.

We next provide details and describe specific choices for the
resonator and system parameters. The {\sc PLL} bandwidth is typically
set to a small fraction of the resonance frequency and is limited by
the capabilities of the loop components such as the {\sc LIA}.  We
choose the controller parameters $K_p$ and $K_i$ as follows, as
suggested in~\cite{olcum2015high},
\begin{equation}
K_p = \omega_{\text{\tiny PLL}} \quad,\quad K_i = \frac{\omega_{\text{\tiny PLL}} }{\tau_r}  
\label{eqn:PIpars}  
\end{equation}
where $\omega_{\text{\tiny PLL}} $ is the desired loop bandwidth.  If
we substitute \eqref{eqn:PIpars} into
\eqref{eqn:plltrfunfromthetntodeltaomega}, a pole-zero cancellation
occurs in the transfer function, as shown in~\cite{olcum2015high},
and simplifies to
\begin{equation}
\begin{aligned}
H_{\theta_n}^{\Delta\Omega}\pqty{s} & = \frac{1}{\tau_r}\bqty{
\frac{ \omega_{\text{\tiny PLL}} \,H_L\pqty{s}}{s + \omega_{\text{\tiny PLL}} \,H_L\pqty{s}}} \\
& = \frac{1}{\tau_r}\bqty{
\frac{H_L\pqty{s}}{H_L\pqty{s} + \frac{s}{\omega_{\text{\tiny PLL}} }}}
\end{aligned}
\label{eqn:plltrfunfromthetntodeltaomegacancel}  
\end{equation}
The bandwidth of the filters $H_L\pqty{s}$ in demodulator are set to be
larger than the desired loop bandwidth $\omega_{\text{\tiny PLL}}$, implying 
\begin{equation}
H_L\pqty{j\omega_{\text{\tiny PLL}} } \approx 1   
\label{eqn:HLwL}  
\end{equation}
Hence, we have
\begin{equation}
\begin{aligned}
H_{\theta_n}^{\Delta\Omega}\pqty{s} \approx \frac{1}{\tau_r}
\left\{ \mqty{ 
\frac{1}{1 + \frac{s}{\omega_{\text{\tiny PLL}} }}&\text{for}\:\vqty{s} \leq \omega_{\text{\tiny PLL}} \\ \\
\frac{\omega_{\text{\tiny PLL}} \,H_L\pqty{s}}{s}&\text{for}\:\vqty{s} \gg \omega_{\text{\tiny PLL}} 
}\right.
\end{aligned}
\label{eqn:plltrfunfromthetntodeltaomegaapprox}  
\end{equation}
Thus, the loop bandwidth is indeed set to be $\omega_{\text{\tiny
    PLL}} $, with an effective one-pole, first-order loop dynamics.   
Resonator and system parameters are chosen as follows, similar to the
choices in~\cite{roy2018improving}. 
{\sc PLL} bandwidth is set as 
\[ \omega_{\text{\tiny PLL}} = 5\!\!\times\!\!10^{-5}\:\omega_o \]
Hence, the loop time-constant is equal to $2\!\!\times\!\!10^{4}$
periods of the high-frequency signal at the output of the resonator.
The low-pass filters in the demodulator are chosen as $4$th order
Butterworth filters with pass-band edge frequency set to
\[ \omega_L = 8\: \omega_{\text{\tiny PLL}} \]
We define the {\it dynamic range} $\text{\small\textsf{DR}}$ for the
resonator in terms of $\text{\small\textsf{SNR}}$:
\begin{equation}
\text{\small\textsf{DR}} = 20\log_{10}\pqty{\text{\small\textsf{SNR}}}  
\label{eqn:defnDR}  
\end{equation}
We present results for two cases, a low quality factor, $Q=50$, and a
high one, $Q=10000$, with the same resonance frequency. In order to
compare these two cases at the onset of Duffing nonlinearity, we
adjust the drive strength for the resonator in such a way so that
\begin{equation}
\text{\small\textsf{SNR}} \propto \frac{1}{Q}
\label{eqn:QvsSNR}  
\end{equation}
as suggested in~\cite{roy2018improving}. For $Q=10000$, we choose
$\text{\small\textsf{DR}} \approx 60\,\text{dB}$. For $Q=50$ then, we
set 
\[ \text{\small\textsf{DR}} \approx 60\,\text{dB} +
20\log_{10}\pqty{\frac{10000}{50}} \approx 106\,\text{dB}
\]
in accordance with \eqref{eqn:QvsSNR}.  
We choose $\text{\small\textsf{BW}}$ in the
$\text{\small\textsf{SNR}}$ definition same as the bandwidth of the
filters in the demodulator
\[
\text{\small\textsf{BW}} = \omega_L = 8\: \omega_{\text{\tiny PLL}}
\] 
We set the duration of the simulation to be $10^{8}$ periods
of the high-frequency signal at the output of the resonator, which is
equal to $5\!\!\times\!\!10^{3}$ loop time constants. 

\begin{figure*}[t]
\centering
\includegraphics[width=0.99\textwidth]{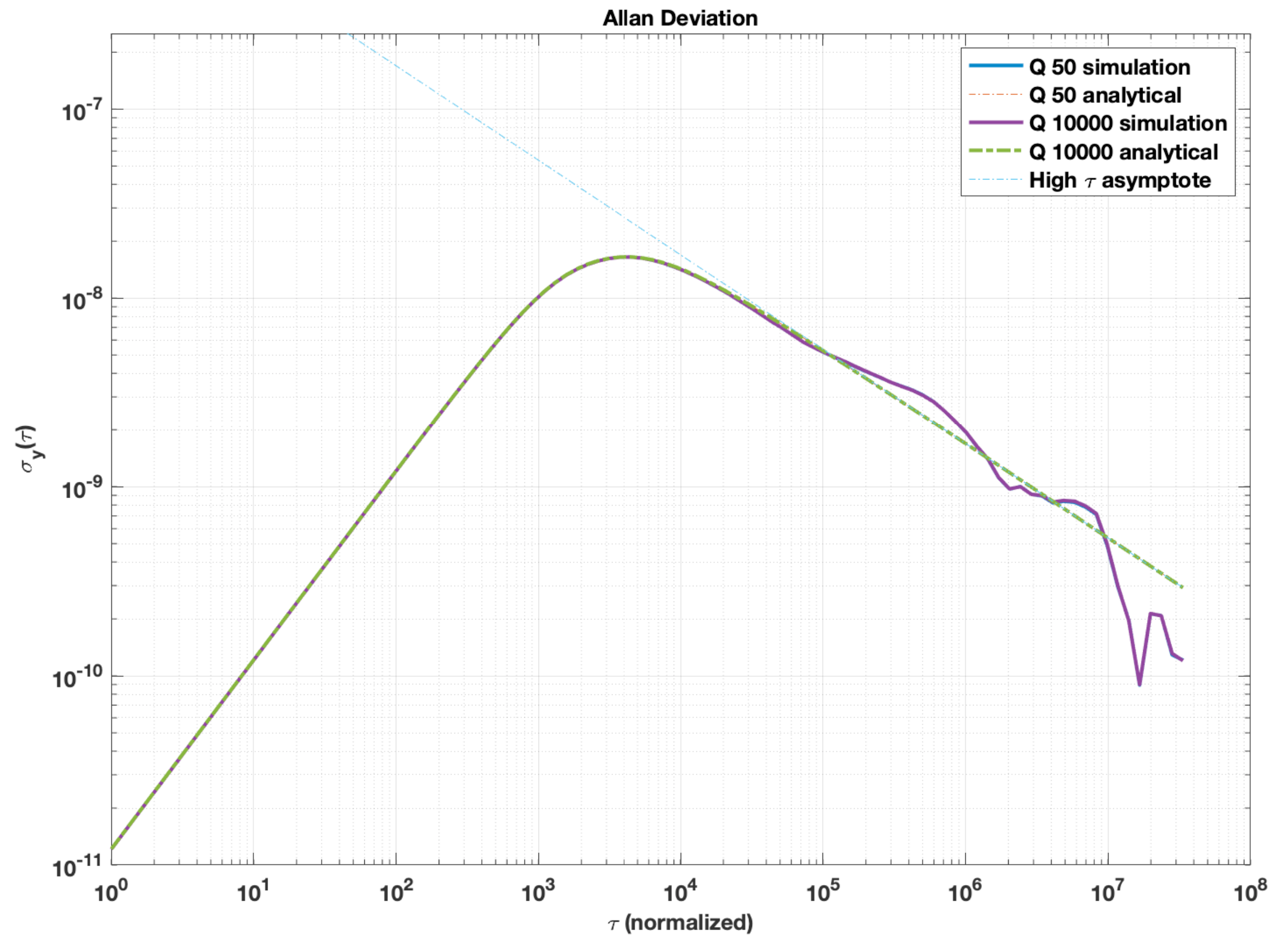}
\caption{Allan Deviation}
\label{fig:allendev}
\end{figure*}

In Figure~\ref{fig:allendev}, we present results obtained for Allan
Deviation, for $Q=50$ and $Q=10000$, based on both the analytical
derivations in Section~\ref{sec:theory} and the simulations. For the
analytical results presented in Figure~\ref{fig:allendev}, the Allan
Variance integral in \eqref{eqn:allanvarfrompsd} was evaluated
numerically. For the results based on simulation, Allan Variance was
estimated from simulation time-series data using the {\it overlapping
  Allan variance estimator}~\cite{wiki:Allanvariance}. The $\tau$ axis in
Figure~\ref{fig:allendev} is normalized, i.e., shows the number of
cycles of the resonator output signal. We note that the high-$\tau$
approximation that was derived in \eqref{eqn:allanvdevlargetauQ}
indeed coincides with the results in Figure~\ref{fig:allendev} for
$\tau>10^{5}$, forming a high-$\tau$ asymptote. The loop time-constant
is $2\!\!\times\!\!10^{4}$ (normalized).

\begin{figure*}[t]
\centering
\includegraphics[width=0.99\textwidth]{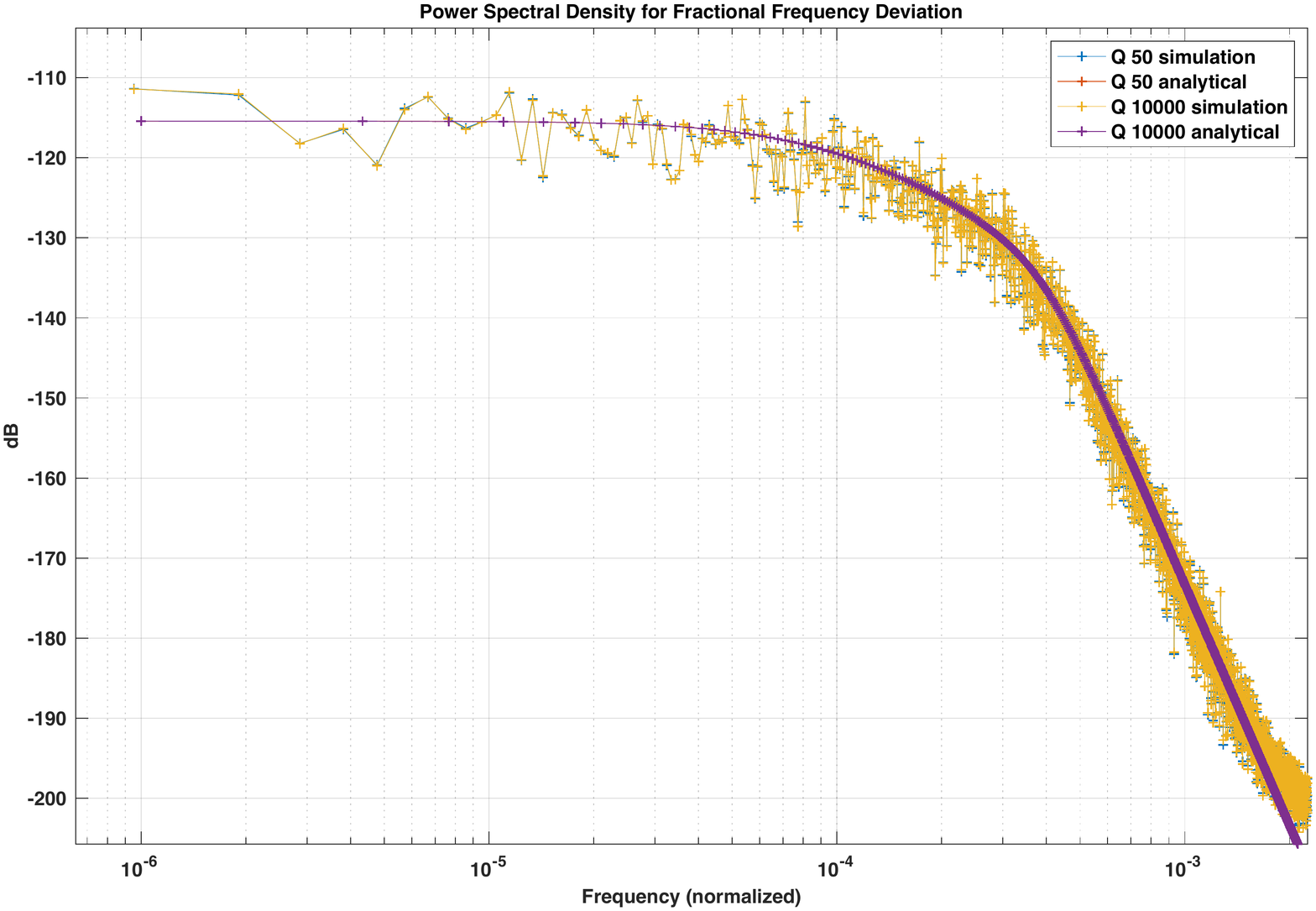}
\caption{{\sc PSD} for Fractional Frequency Deviation}
\label{fig:psdfracfreq} 
\end{figure*}

In Figure~\ref{fig:psdfracfreq}, we present results obtained for the
{\sc PSD} of fractional frequency deviation $y\pqty{t}$ defined by
\eqref{eqn:fracfreqdev}. This figure contains results for $Q=50$ and
$Q=10000$, based on both the analytical derivations in
Section~\ref{sec:theory} and the simulations. The analytical results
presented in Figure~\ref{fig:psdfracfreq} were obtained by simply
evaluating \eqref{eqn:psdy}, \eqref{eqn:psdphaseandfreqdev} and
\eqref{eqn:plltrfunfromthetntodeltaomega}.  The results based on
simulations were obtained via spectral estimation from simulation
time-series data using {\it Welch's method}~\cite{welch_use_1967}. The
frequency axis in Figure~\ref{fig:psdfracfreq} is normalized with the
resonance frequency. We note that the {\sc PSD} of frequency deviation
has a Lorentzian shape, for low frequencies and up to and exceeding
the loop bandwidth, as predicted by
\eqref{eqn:plltrfunfromthetntodeltaomegaapprox}. For larger
frequencies on the other hand, {\sc PSD} exhibits a faster roll-off
due to the effect of the high-order low-pass filters in the
demodulator.

We note that there is excellent agreement between the analytical
results and the ones obtained from simulations, both for Allan
Deviation and the spectrum of the frequency deviation. Discrepancies
at larger values of $\tau$ are expected for Allan Deviation, due to
the inaccuracy of Allan Variance estimation for large values of $\tau$
from time-limited simulation data. The agreement at low values of $\tau$, below
and exceeding the loop time constant, is excellent. 

\begin{figure*}[t]
\centering
\includegraphics[width=0.99\textwidth]{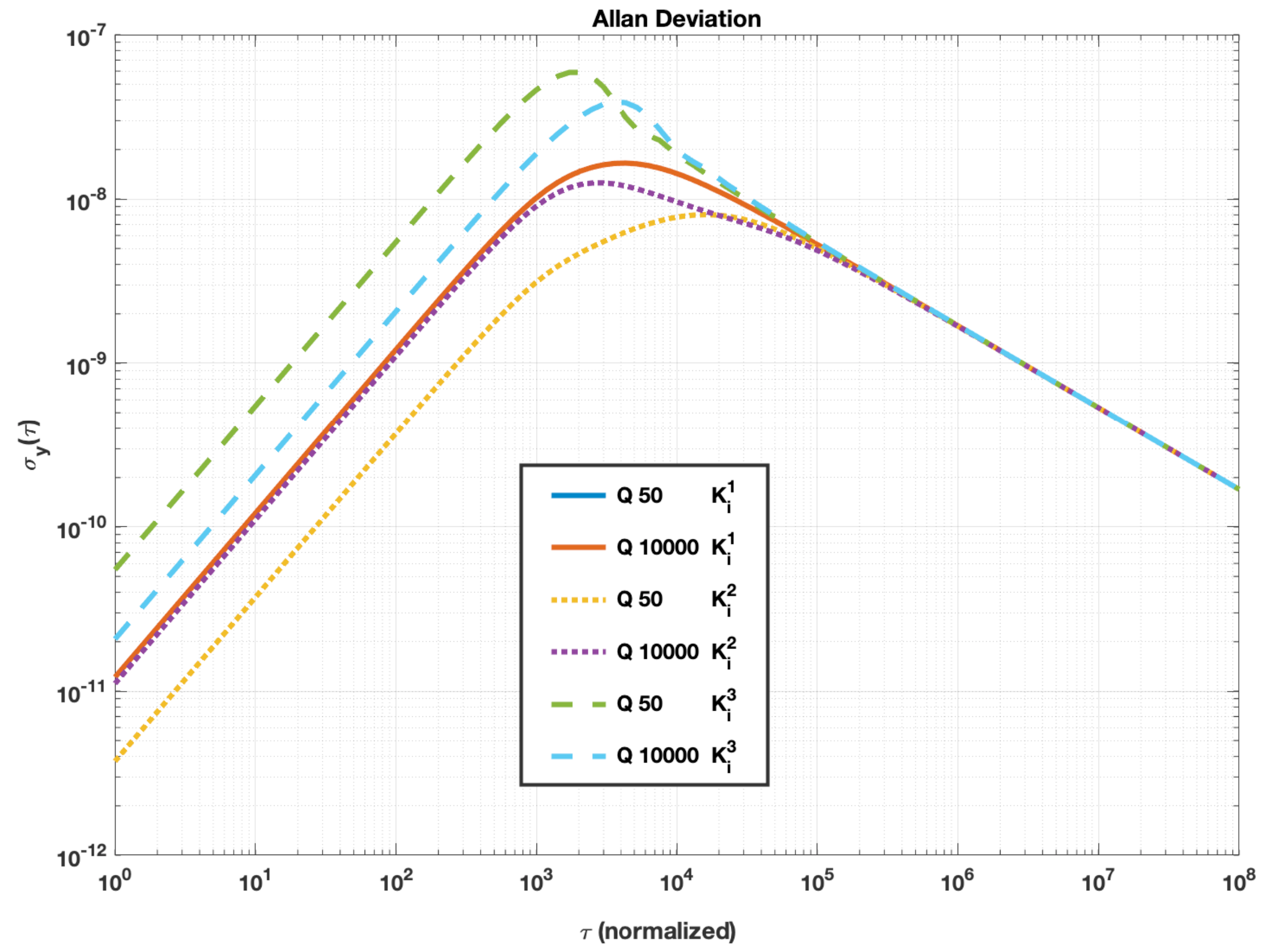}
\caption{Allan Deviation versus Controller Parameter $K_i$}
\label{fig:allendevKi}
\end{figure*}

\begin{figure*}[t]
\centering
\includegraphics[width=0.99\textwidth]{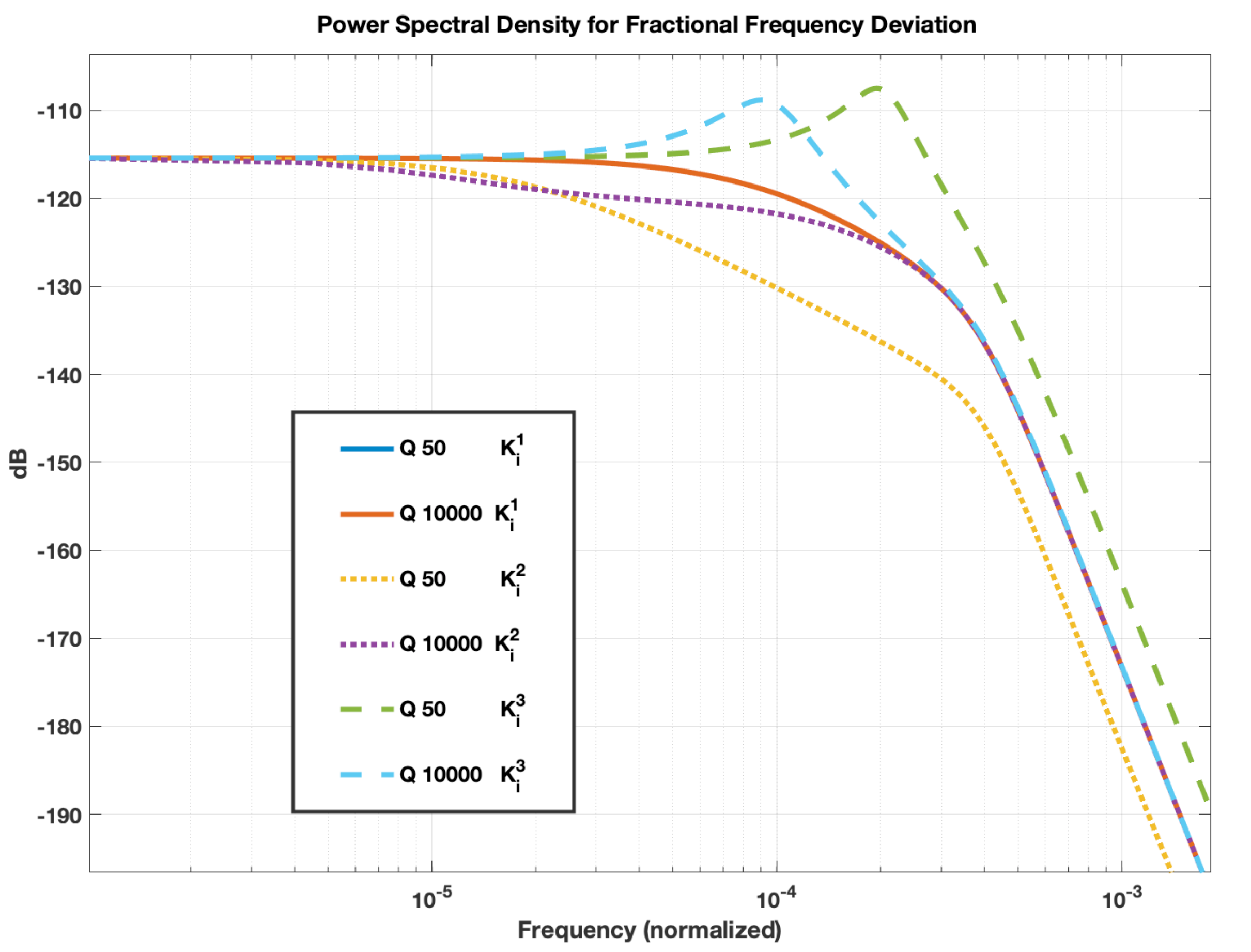}
\caption{{\sc PSD} for Fractional Frequency Deviation versus Controller Parameter $K_i$}
\label{fig:psdfracfreqKi}
\end{figure*}

It is noteworthy that the simulation results perfectly agree with the
analytical results derived in Section~\ref{sec:theory}: If
$\text{\small\textsf{SNR}}$ and $Q$ are related as in
\eqref{eqn:QvsSNR}, the Allan Deviation is independent of $Q$ for all
values of $\tau$. The results presented in Figure~\ref{fig:allendev}
and Figure~\ref{fig:psdfracfreq} for the two $Q$ values fall on top of
each other. This correspondence is exact, due to the scaling factor in
\eqref{eqn:allanvdevlargetauQ} and the choice for the controller
parameters in \eqref{eqn:PIpars}. This particular choice for the
controller parameters $K_p$ and $K_i$ in \eqref{eqn:PIpars} is unique
in the sense that it results in a pole-zero
cancellation~\cite{olcum2015high} in the transfer function
$H_{\theta_n}^{\Delta\Omega}\pqty{s}$, making its poles and zeros
independent of $Q$ or the resonator time constant $\tau_r$.  (We note
that, in this case, the poles/zeros are independent of $\tau_r$, but
$\tau_r$ still appears in a front factor in
$H_{\theta_n}^{\Delta\Omega}\pqty{s}$.)  Thus, Allan Deviation is then
independent of $Q$ for all values of $\tau$, when $\text{\small
  \textsf{SNR}}$ is adjusted so as to hold the product
$Q\,\text{\small \textsf{SNR}}$ constant. On the other hand, if the
controller parameters $K_p$ and $K_i$ are not chosen as in
\eqref{eqn:PIpars}, or if another type of controller is used, then the
Allan Deviation will {\em not} be independent of $Q$ for all values of
$\tau$, even when $Q\,\text{\small \textsf{SNR}}$ is held
constant. However, we emphasize, the high-$\tau$ asymptote (for values
of $\tau$ larger than the loop time constant) will always be given by
\eqref{eqn:allanvdevlargetauQ}, i.e., independent of $Q$ with constant
$Q\,\text{\small \textsf{SNR}}$. The controller parameters and the
particular controller design has an effect on the Allan Deviation only
for low values of $\tau$, at or below the loop time constant, which is
not significant from a practical point of view, since the {\sc PLL}
system is useful in tracking frequency deviations only at time scales
longer than the loop time constant. In order to illustrate this, we
show in Figure~\ref{fig:allendevKi}, the Allan Deviation (based on the
theory presented in the paper) for three different choices for $K_i$:
\begin{equation}
K_i^1 = \frac{\omega_{\text{\tiny PLL}} }{\tau_r},\quad K_i^2 = \frac{\omega_{\text{\tiny PLL}} }{3\,\tau_r},\quad K_i^3 = \frac{3\,\omega_{\text{\tiny PLL}} }{\tau_r}  
\label{eqn:PIparsKi}  
\end{equation}
Above, $K_i^1$ is the same as in \eqref{eqn:PIpars}, and all other
system and resonator parameters were chosen as described before.  We
observe in Figure~\ref{fig:allendevKi} that, for $K_i^2$, lower $Q$
yields seemingly better performance for low values of $\tau$. This is
due to the fact that the transfer
$H_{\theta_n}^{\Delta\Omega}\pqty{s}$ function has a smaller effective
bandwidth for the particular placement of its poles and zeros for the
lower $Q$ case.  {\bf However, this also means that the {PLL} will be
  attenuating the frequency shifts induced by events of interest,
  e.g., addition of mass, more severely,} rendering it not useful for
sensing at time scales where the Allan Deviation for lower $Q$ is
smaller than the case for higher $Q$. This can be observed in
Figure~\ref{fig:psdfracfreqKi} (based on the theory presented in the
paper), where {\sc PSD} of fractional frequency deviation is shown for
both $Q=50$ and $Q=10000$, for the three values of $K_i$ in
\eqref{eqn:PIparsKi}.  We observe in Figures~\ref{fig:allendevKi} and
\ref{fig:psdfracfreqKi} that the lower Allan Deviation is accompanied
with more severe attenuation of frequency deviation.  With the {\sc
  PLL} architecture considered, one can not improve performance at
practically relevant time scales (corresponding to the high-$\tau$
asymptote in Figure~\ref{fig:allendevKi} where all curves coincide) by
simply optimizing the controller parameters.  The controller parameter
choice in \eqref{eqn:PIpars} in fact strikes a good balance between
Allan Deviation and the attenuation of frequency deviations. Since the
frequency deviation {\sc PSD} in this case is maximally
flat~\cite{olcum2015high} below the low bandwidth, the step response
of the {\sc PLL} system will not exhibit any ringing and overshoots.

The results we have derived and reported here are in stark contrast to
the theory and results presented in~\cite{roy2018improving}. The
theory presented in~\cite{roy2018improving} does not
consider the closed-loop dynamics of the {\sc PLL} tracking system.
The flattening of the phase spectrum at low frequencies is the basis
of the claim in~\cite{roy2018improving} that the sensor performance
can be improved with larger damping.  Our theory and results show that
there is no such flattening of the phase noise spectrum under feedback
in a {\sc PLL}. In fact, as phase deviation $\theta_o\pqty{t}$ is
simply the integral of the frequency deviation $\Delta\Omega\pqty{t}$,
and since $\Delta\Omega\pqty{t}$ has a Lorentzian {\sc PSD}, the phase
spectrum (given by $1/\omega^2$ times the spectrum of
$\Delta\Omega\pqty{t}$) does not flatten at low frequencies. On the
contrary, the phase spectrum keeps increasing as frequency is lowered,
a signature of nonstationary and random walk phase noise.
In~\cite{roy2018improving}, it is suggested that one can circumvent
random walk phase noise in a {\sc PLL} based system if a high
precision {\sc NCO/VCO} is used. Our theory suggests otherwise.  A
high precision {\sc NCO/VCO} will not produce (or produce very little)
random walk phase noise arising from its own, internal noise sources,
{\it provided that} it is controlled with a constant, noiseless
frequency control input. However, in the context of a {\sc PLL}, the
frequency control input is noisy, due to unavoidable noise from other
sources (resonator, amplifiers, etc.) circulating around the loop and
shaped by the loop dynamics. This is a fundamental aspect of {\sc PLL}
operation.


%% file: conclusion.tex
\section{Conclusions}
\label{sec:conc}
We have presented a theory and noise analysis framework, and an
associated simulator, for {\sc PLL} based resonant sensors, which is
useful in deciphering the fundamental limitations and understanding
basic trade-offs due to inherent noise and fluctuations arising from a
number of sources. The framework we have described enables a firm
analytical handle on the problem, but without forfeiting rigor and
precision. In this paper, we considered a setting where the dominant
source of noise is the thermo-mechanical noise of the resonator. In
future work, we will extend the analysis framework and the simulator
to take into account other types of noise and nonideal dynamics in the
resonator~\cite{cleland2002noise}, electronic amplifier and
instrumentation noise, fluctuations that arise from the actuation and
sensing mechanisms in the mechanical, electrical and optical domains,
non-negligible phase noise of the controlled oscillator (signal
generator), quantization noise in digital (DSP/FPGA) realizations of
some {\sc PLL} loop components. Furthermore, we will also develop
extensions so that the analysis framework can be applied to a variety
of {\sc PLL} based sensor configurations, such as multi-mode mass
spectrometry with multiple {\sc PLL}s~\cite{hanay2012single}, and
nonlinear nano-mechanical trajectory-locked loop ({\sc
  TLL})~\cite{yuksel2019nonlinear} based sensing. Even though we have
shown that lowering the quality factor of the resonator does not
result in the claimed performance improvement, one may be able to
obtain better performance by optimizing the controller and the filters
in the demodulator in the presence of a variety of noise sources, that
we plan to investigate in the near future using the proposed analysis
framework.

%% file: noisecalibration.tex
\section{Thermo-mechanical Noise}
\label{sec:thermechnoiseintegral}
In this appendix, we evaluate the integral in
\eqref{eqn:thmnoiseintegral}, which is reproduced here for
convenience:
\begin{equation}
E_{K} = \frac{1}{2}m\,\frac{1}{2\pi} \int_{-\infty}^{\infty}
\omega^2\;\vb{S}_x\pqty{\omega}\,\dd \omega
\label{eqn:thmnoiseintegralapx} 
\end{equation} 
where 
\begin{equation}
\vb{S}_x\pqty{\omega} = \vqty{H_{R}^{BP}\pqty{j\,\omega}}^2\;S_{thm}\pqty{\omega}
\label{eqn:sxdefn}
\end{equation} 
with
\begin{equation}
H_{R}^{BP}\pqty{s}=\frac{X\pqty{s}}{F\pqty{s}}=\frac{1}{m}\,\frac{1}{s^2+\Gamma\,s+\omega_r^2}
\label{eqn:resonatortranfunapx}
\end{equation}
and 
\begin{equation}
S_{thm}\pqty{\omega} = 2\,m\,\Gamma\,k_{B}\,T
\label{eqn:thmnoisePSDapx} 
\end{equation}
Please see~\cite{karr1953effective} where a
power-law noise source driving a parallel {\sc RLC} circuit is considered. 
The below treatment is based on~\cite{karr1953effective}, with
\eqref{eqn:thmnoiseintegralapx} as a
special case of the problem considered there. 

We substitute \eqref{eqn:thmnoisePSDapx},
\eqref{eqn:resonatortranfunapx} and \eqref{eqn:sxdefn} in
\eqref{eqn:thmnoiseintegralapx}, and manipulate to simplify and obtain 
\begin{equation}
E_{K} = \frac{\Gamma\,k_{B}\,T}{2\pi} 
\int_{-\infty}^{\infty}  
\frac{\omega^2}{\pqty{\omega^2-\omega_r^2}^2+\Gamma^2\,\omega^2}
\dd \omega
\label{eqn:thmnoiseintegralapx2} 
\end{equation} 
We substitute \eqref{eqn:gammadefn} into the above and reorganize:
\begin{equation}
\begin{aligned}
E_{K} & = \frac{\omega_r\,k_{B}\,T}{2\pi\,Q} 
\int_{-\infty}^{\infty}  
\frac{\omega^2}{\pqty{\omega^2-\omega_r^2}^2+\pqty{\frac{\omega_r\,\omega}{Q}}^2}
\dd \omega\\
& = \frac{k_{B}\,T}{2\pi\,Q\,\omega_r} 
\int_{-\infty}^{\infty}  
\frac{1}{\pqty{\frac{\omega^2-\omega_r^2}{\omega_r\,\omega}}^2+\frac{1}{Q^2}}
\dd \omega\\
& = \frac{k_{B}\,T}{2\pi\,Q\,\omega_r} 
\int_{-\infty}^{\infty}  
\frac{1}{\pqty{\frac{\omega}{\omega_r}-\frac{\omega_r}{\omega}}^2+\frac{1}{Q^2}}
\dd \omega
\end{aligned}
\label{eqn:thmnoiseintegralapx3} 
\end{equation} 
Next, a new variable of integration is defined as in~\cite{karr1953effective} 
\begin{equation}
\beta = \pqty{\frac{\omega}{\omega_r}}^2,\quad \dd \beta =
2\frac{\omega}{\omega_r}\frac{\dd \omega}{\omega_r},\quad \dd \omega =
\frac{\omega_r}{2}\frac{\dd \beta}{\sqrt{\beta}}
\end{equation}
We rewrite the integral in \eqref{eqn:thmnoiseintegralapx3} as follows:
\begin{equation}
\begin{aligned}
E_{K}  & = \frac{k_{B}\,T}{2\pi\,Q} 
\int_{0}^{\infty}  
\frac{1}{\pqty{\sqrt{\beta}-\frac{1}{\sqrt{\beta}}}^2+\frac{1}{Q^2}}
\frac{\dd \beta}{\sqrt{\beta}} \\
& =  \frac{k_{B}\,T}{2\pi\,Q}
\int_{0}^{\infty}  
\frac{\sqrt{\beta}}{\beta^2+2\,\beta\pqty{\frac{1}{2\,Q^2}-1}+1}
\dd \beta \\
& =  \frac{k_{B}\,T}{2\pi}
\int_0^{\infty}  
\frac{1}{Q}\frac{\sqrt{\beta}}{\beta^2+2\,\beta\pqty{\frac{1}{2\,Q^2}-1}+1}
\dd \beta
\end{aligned}
\label{eqn:thmnoiseintegralapx4} 
\end{equation} 
where we used the fact that 
\[ \beta = \pqty{\frac{\omega}{\omega_r}}^2 \geq 0\quad\text{for}\:
-\infty < \omega < +\infty
\]
and assumed 
\[ 
0 < \frac{1}{2\,Q^2}-1 < 1, \quad\text{equivalently,}\quad Q > \frac{1}{2} 
\]
Based on the above, the double-sided integral in
\eqref{eqn:thmnoiseintegralapx3} was turned into a one-sided one with
the new integration variable $\beta$, since the integrand is an even
function of $\omega$.  The final form of the integral in
\eqref{eqn:thmnoiseintegralapx4} above can be evaluated, by defining~\cite{karr1953effective} 
\[
\cos\lambda = \frac{1}{2\,Q^2}-1
\]
and using contour integration in the complex
plane~\cite{karr1953effective} to yield $\pi$. 
Please see~\cite{karr1953effective}.

Finally, we have
\begin{equation}
\begin{aligned}
E_{K}  =  \frac{k_{B}\,T}{2\pi}\:\pi = \frac{k_{B}\,T}{2}
\end{aligned}
\label{eqn:thmnoiseintegralapx5} 
\end{equation}

%% file: spectrumcyclo.tex
\section{Spectral Characterizations and Filtering of\\
  Cyclo-Stationary Processes} 
\label{sec:specPSDcyclo}
Please see~\cite{GardnerRP,RoLoFe97jssc,demir1998Kluwer} for details
regarding the spectral characterization and filtering of
cyclo-stationary random processes. Our treatment below is based
on~\cite{GardnerRP,RoLoFe97jssc,demir1998Kluwer}.

Let $s\pqty{t}$ be a zero-mean, stationary Gaussian random process
with the auto-correlation function
\begin{equation}
R_s\pqty{\eta} = \vb{E}\bqty{s\pqty{t+\eta/2}\,s\pqty{t-\eta/2}}
\label{eqn:autocorsta}  
\end{equation}
where $ \vb{E}\bqty{\cdot}$ denotes the probabilistic expectation
operator.
$R\pqty{\eta}$ is a function of only $\eta$, not $t$, due to the
stationarity of the process. 
The {\sc PSD} of $s\pqty{t}$ is defined as the Fourier transform of
$R_s\pqty{\eta}$
\begin{equation}
S_s\pqty{\omega} = {\cal F}\Bqty{R_s\pqty{\eta}}
\label{eqn:psdsta}  
\end{equation}
Let $m\pqty{t} = A_o\cos\pqty{\omega_o t}$ be a periodic modulating
signal. We obtain the modulated signal (random process) $c\pqty{t}$ from $s\pqty{t}$ as
follows 
\begin{equation}
c\pqty{t} = m\pqty{t}\,s\pqty{t} = A_o\cos\pqty{\omega_o t}\,s\pqty{t}
\label{eqn:modulated}  
\end{equation}
The auto-correlation function of $c\pqty{t}$ is also a (periodic)
function of $t$ and can be computed as follows
\begin{equation}
\begin{aligned}
R_c\pqty{t,\eta} &= \vb{E}\bqty{c\pqty{t+\eta/2}\,c\pqty{t-\eta/2}} \\
&=\vb{E}\bqty{m\pqty{t+\eta/2}\,m\pqty{t-\eta/2}\:s\pqty{t+\eta/2}\,s\pqty{t-\eta/2}}\\
&=m\pqty{t+\eta/2}\,m\pqty{t-\eta/2}\:\vb{E}\bqty{s\pqty{t+\eta/2}\,s\pqty{t-\eta/2}}\\
&=m\pqty{t+\eta/2}\,m\pqty{t-\eta/2}\:R_s\pqty{\eta}
\end{aligned}
\label{eqn:autocorcyc}  
\end{equation}
where 
\begin{equation}
\begin{aligned}
m&\pqty{t+\eta/2}\,m\pqty{t-\eta/2} \\ = A_o^2&\cos\pqty{\omega_o
  \pqty{t+\eta/2}}\cos\pqty{\omega_o \pqty{t-\eta/2}}\\
=\frac{A_o^2}{4}&
  \bqty{e^{j\pqty{\omega_o
  \pqty{t+\eta/2}}}+e^{-j\pqty{\omega_o
  \pqty{t+\eta/2}}}}\\
  & \bqty{e^{j\pqty{\omega_o
  \pqty{t-\eta/2}}}+e^{-j\pqty{\omega_o
  \pqty{t-\eta/2}}}}\\
=\frac{A_o^2}{4}&\bqty{e^{j\omega_o\eta} + e^{-j\omega_o\eta}+ e^{j2\omega_o t} + e^{-j2\omega_o t} }
\end{aligned}
\label{eqn:modusignalautocor}  
\end{equation}
The {\sc PSD} of $c\pqty{t}$ is also a (periodic) function of $t$, in
addition to $\omega$. The $t$ dependence can be expanded into a
Fourier series~\cite{GardnerRP,RoLoFe97jssc,demir1998Kluwer}
\begin{equation}
\begin{aligned}
S_c\pqty{t,\omega} = {\cal F}\Bqty{R_c\pqty{t,\eta}}
= \sum_{k} S_c^{\pqty{k}}\pqty{\omega}\,e^{j k\omega_o t}
\end{aligned}
\label{eqn:psdcycdef}  
\end{equation}
where $S_c^{\pqty{k}}\pqty{\omega}$ are called the {\it cyclic
  spectra}. In \eqref{eqn:psdcycdef}, the Fourier transform ${\cal
  F}\Bqty{R_c\pqty{t,\eta}}$ is with respect to the variable $\eta$.
For a stationary process, we have $S_c^{\pqty{k}}\pqty{\omega}=0$ for
$k>0$. In this case, $S_c^{\pqty{0}}\pqty{\omega}$ corresponds to the
usual {\sc PSD} for a stationary
process~\cite{GardnerRP,RoLoFe97jssc,demir1998Kluwer}.

Using \eqref{eqn:autocorcyc}, \eqref{eqn:modusignalautocor} and
\eqref{eqn:psdcycdef}, we obtain
\begin{equation}
\begin{aligned}
S_{c}^{\pqty{0}}\pqty{\omega} &=
\frac{A_o^2}{4}\bqty{S_{s}\pqty{\omega-\omega_o}+S_{s}\pqty{\omega+\omega_o}}\\
 S_{c}^{\pqty{2}}\pqty{\omega} & =  S_{c}^{\pqty{-2}}\pqty{\omega} =
\frac{A_o^2}{4}S_{s}\pqty{\omega}\\
 S_{c}^{\pqty{k}}\pqty{\omega} &= 0\quad \text{for all other}\;\;k
\end{aligned}
\label{eqn:psdcyc}  
\end{equation}
Next, we consider the (low-pass) filtering of $c\pqty{t}$ with a
(linear and time-invariant) filter frequency response
$H_L\pqty{j\omega}$~\cite{GardnerRP,RoLoFe97jssc}. The output of the filter,
denoted by $c_L\pqty{t}$, is in general also a cyclo-stationary
process. It can be shown that~\cite[eqn. 2.139]{demir1998Kluwer} the
cyclic spectra of $c_L\pqty{t}$ can be computed with
\begin{equation}
S_{cL}^{\pqty{k}}\pqty{\omega} = H_L\pqty{j\omega+j\tfrac{k\,\omega_o}{2}}\,S_{c}^{\pqty{k}}\pqty{\omega}\,H_L^*\pqty{j\omega-j\tfrac{k\,\omega_o}{2}}  
\label{eqn:cycthrufilter}  
\end{equation}
where $\cdot^*$ denotes the complex-conjugate. The input cyclic
spectra $S_{c}^{\pqty{k}}\pqty{\omega}$ is nonzero only for
$k=0,\pm 2$ for which we use \eqref{eqn:cycthrufilter} to obtain  
\begin{equation}
\begin{aligned}
S_{cL}^{\pqty{0}}\pqty{\omega} &= \vqty{H_L\pqty{j\omega}}^2\,S_{c}^{\pqty{0}}\pqty{\omega}
\\
 S_{cL}^{\pqty{2}}\pqty{\omega} & =
H_L\pqty{j\pqty{\omega+\omega_o}}\,S_{c}^{\pqty{2}}\pqty{\omega}\,H_L^*\pqty{j\pqty{\omega-\omega_o}}
\\
S_{cL}^{\pqty{-2}}\pqty{\omega} & = H_L\pqty{j\pqty{\omega-\omega_o}}\,S_{c}^{\pqty{-2}}\pqty{\omega}\,H_L^*\pqty{j\pqty{\omega+\omega_o}}
\\
 S_{cL}^{\pqty{k}}\pqty{\omega} &= 0\quad \text{for all other}\;\;k
\end{aligned}
\label{eqn:psdcycthrufilterspec}  
\end{equation}
The result above is valid for any input stationary process $s\pqty{t}$
and for any (linear and time-invariant) filter $H_L\pqty{j\omega}$. 

We next consider the case when $H_L\pqty{j\omega}$ is a low-pass
filter with an effective bandwidth that is much less than $\omega_o$,
satisfying $H_L\pqty{\pm j\omega_o} \approx H_L\pqty{\pm j 2\omega_o}
\approx 0$. This implies that
\begin{equation}
\mqty{H_L\pqty{j\pqty{\omega-\omega_o}}\\\times H_L^*\pqty{j\pqty{\omega+\omega_o}}}
  = \left\{
    \mqty{H_L\pqty{-j\omega_o}\,H_L^*\pqty{j\omega_o}\approx 0
      & \omega \approx 0 \\
H_L\pqty{0}\,H_L^*\pqty{j 2\omega_o}\approx 0
      & \omega \approx \omega_o \\
H_L\pqty{-j 2\omega_o}\,H_L^*\pqty{0}\approx 0
      & \omega \approx -\omega_o
} \right.  
\label{eqn:lowpassapprox}  
\end{equation}
In fact, we have 
\begin{equation}
H_L\pqty{j\pqty{\omega-\omega_o}}\,H_L^*\pqty{j\pqty{\omega+\omega_o}}
\approx 0 \quad \text{for all}\;\;\omega 
\label{eqn:lowpassaprox2}  
\end{equation}
Then, based on \eqref{eqn:psdcycthrufilterspec} and
\eqref{eqn:lowpassaprox2}, we conclude 
\begin{equation}
 S_{cL}^{\pqty{k}}\pqty{\omega} = 0\quad \text{for}\;\;k>0  
\label{eqn:higordcycspec}  
\end{equation}
That is, the output of the low-pass filter $H_L\pqty{j\omega}$ becomes
a stationary process with {\sc PSD}
\begin{equation}
\begin{aligned}
S_{cL}\pqty{\omega} &= S_{cL}^{\pqty{0}}\pqty{\omega} =
\vqty{H_L\pqty{j\omega}}^2\,S_{c}^{\pqty{0}}\pqty{\omega}\\
&= \frac{A_o^2}{4}\,\vqty{H_L\pqty{j\omega}}^2\,\bqty{S_{s}\pqty{\omega-\omega_o}+S_{s}\pqty{\omega+\omega_o}}
\end{aligned}
\label{eqn:statpsdfilterout}  
\end{equation}
Thus, the low-pass filter {\it stationarizes} the cyclo-stationary
noise process $c\pqty{t}$ by removing the high-order cyclic
components~\cite{RoLoFe97jssc}. Furthermore,
\eqref{eqn:statpsdfilterout} reveals that there is {\it noise folding}
in the frequency domain due to the modulation in
\eqref{eqn:modulated}~\cite{RoLoFe97jssc}. That is, the noise
components of $s\pqty{t}$ at frequencies $\omega-\omega_o$ and
$\omega+\omega_o$ fold and both generate a noise component at $\omega$
in $c_L\pqty{t}$. Furthermore, the low-pass filter $H_L\pqty{j\omega}$
removes any high-frequency noise components in $c_L\pqty{t}$,
producing a low-pass noise {\sc PSD}.

%% file: allandev.tex
\section{Allan Deviation} 
\label{sec:allendev}
In this appendix, we derive 
\eqref{eqn:allanvarfrompsd}, which is reproduced here for convenience: 
\begin{equation}
\sigma_y^2\pqty{\tau} = \frac{4}{\pi\tau^2}\:\int_{-\infty}^{+\infty}
\frac{\bqty{\sin\pqty{\frac{\omega\,\tau}{2}}}^4}{\omega^2}\:S_y\pqty{\omega}\,\dd \omega  
\label{eqn:allanvarfrompsdapx}  
\end{equation}
Please
see~\cite{allan1966statistics,rubiola2009phase,wiki:Allanvariance} for
details on Allan Variance. Our treatment below is based
on~\cite{allan1966statistics,rubiola2009phase,wiki:Allanvariance}.

We define the timing deviation $\alpha\pqty{t}$ as the integral of the
fractional frequency deviation $y\pqty{t}$:
\begin{equation}
y\pqty{t} = \dv{}{t}\alpha\pqty{t}
\label{eqn:alphadefn}  
\end{equation}
Thus, the averaged fractional frequency deviation
$\bar{y}\pqty{t,\tau}$, defined by \eqref{eqn:avfracfreqdev}, can be
computed based on $\alpha\pqty{t}$
\begin{equation}
\bar{y}\pqty{t,\tau} = \frac{\alpha\pqty{t+\tau}-\alpha\pqty{t}}{\tau}
\label{eqn:avfracfreqdevalpha}  
\end{equation}
Similarly, samples of $\bar{y}\pqty{t,\tau}$ with a sampling interval
of $\tau$ can be computed with
\begin{equation}
\bar{y}_i =
\frac{\alpha\pqty{\pqty{i+1}\tau}-\alpha\pqty{i\tau}}{\tau} = \frac{\alpha_{i+1}-\alpha_i}{\tau}
\label{eqn:avfracfreqdevalphasample}  
\end{equation}
where we defined
\begin{equation}
\alpha_i = \alpha\pqty{i\tau}
\label{eqn:alphasampleddefn}  
\end{equation}
Then,
\begin{equation}
\begin{aligned}
\sigma_y^2\pqty{\tau} & =
\frac{1}{2}\:\vb{E}\bqty{\pqty{\bar{y}_{i+1}-\bar{y}_i}^2}\\
& =
\frac{1}{2\tau^2}\:\vb{E}\bqty{\pqty{\alpha_{i+2}-2\,\alpha_{i+1}+\alpha_{i}}^2}
\\
& =
\frac{1}{2\tau^2}\:\vb{E}\bqty{\pqty{\alpha\pqty{t+2\tau}-2\,\alpha\pqty{t+\tau}+\alpha\pqty{t}}^2}
\end{aligned}
\label{eqn:allanvaralpha}  
\end{equation}
$\sigma_y^2\pqty{\tau}$ is postulated to be independent of the
sampling times represented by the index $i$. In the above, the $i$th sampling
time $i\tau$ was replaced with $t$.  
We define 
\begin{equation}
\begin{aligned}
\gamma\pqty{t} & =
\alpha\pqty{t+2\tau}-2\,\alpha\pqty{t+\tau}+\alpha\pqty{t} \\ 
& = \bqty{\alpha\pqty{t+2\tau}-\alpha\pqty{t+\tau}} - \bqty{\alpha\pqty{t+\tau}-\alpha\pqty{t}}
\end{aligned}
\label{eqn:allanvarbeta}  
\end{equation}
We note that $\gamma\pqty{t}$ is assumed to be a (wide-sense) stationary
process. However, $\alpha\pqty{t}$ does not need to be, in fact, often it
is not. The stationarity of $\gamma\pqty{t}$ implies that the
expectation in \eqref{eqn:allanvaralpha} is independent of $t$. 
We then have 
\begin{equation}
\sigma_y^2\pqty{\tau} = \frac{1}{2\tau^2}\:\frac{1}{2\pi}\:\int_{-\infty}^{+\infty}
S_\gamma\pqty{\omega}\,\dd \omega  
\label{eqn:allanvarfrompsdbeta}  
\end{equation}
where $S_\gamma\pqty{\omega}$ is the {\sc PSD} of $\gamma\pqty{t}$.

$\gamma\pqty{t}$ is the output of a system that is the cascade
of an integrator and two delay-difference operators, with input set to
$y\pqty{t}$.
In the frequency domain:
\begin{equation}
\begin{aligned}
\frac{\gamma\pqty{s}}{y\pqty{s}} = H_y^\gamma\pqty{s} = \frac{1}{s}\:\pqty{e^{s\tau}-1}^2
\end{aligned}
\label{eqn:allanvarbetatranfun}  
\end{equation}
Hence, we can derive the following using Euler's formula and
trigonometric identities:
\begin{equation}
S_\gamma\pqty{\omega} =   \vqty{H_y^\gamma\pqty{j\omega}}^2
S_y\pqty{\omega} = \frac{16\,\bqty{\sin\pqty{\frac{\omega\tau}{2}}}^4}{\omega^2}\:S_y\pqty{\omega}
\label{eqn:psdgamma}  
\end{equation}
If we substitute \eqref{eqn:psdgamma}
into \eqref{eqn:allanvarfrompsdbeta}, we finally get  
\begin{equation}
\sigma_y^2\pqty{\tau} = \frac{4}{\pi\tau^2}\:\int_{-\infty}^{+\infty}
\frac{\bqty{\sin\pqty{\frac{\omega\,\tau}{2}}}^4}{\omega^2}\:S_y\pqty{\omega}\,\dd \omega  
\label{eqn:allanvarfrompsdapxfinal}  
\end{equation}
Finally, we consider an important special case, where fractional
frequency noise $y\pqty{t}$ is a white Gaussian random process, and
hence the timing deviation $\alpha\pqty{t}$ as its integral is a
Wiener process (Brownian motion). For this case, we can evaluate the
expectation in \eqref{eqn:allanvaralpha} directly, without the need to
evaluate the integral in \eqref{eqn:allanvarfrompsdapxfinal}, using
the following properties of the Wiener process
\begin{equation}
\begin{aligned}
\vb{E}\bqty{\pqty{\alpha\pqty{t}}^2} =
\vb{E}\bqty{\alpha\pqty{t}\,\alpha\pqty{t+\tau}} = c\,t\\
\vb{E}\bqty{
    \pqty{\alpha\pqty{t+2\tau}-\alpha\pqty{t+\tau}}^2} = c\,\tau\\
\vb{E}\bqty{ 
    \pqty{\alpha\pqty{t+\tau}-\alpha\pqty{t}}^2  } = c\,\tau
\end{aligned}
\label{eqn:wienerprop}  
\end{equation}
for some constant $c$ and $\tau\geq 0$. The above follows from the
fact that the Wiener process is the integral of stationary white
Gaussian noise. It has {\it independent increments}~\cite{Grimmet92},
that is, $\alpha\pqty{t+\tau}-\alpha\pqty{t}$ is independent of
$\alpha\pqty{t}$ for $\tau\geq 0$.  Then,
\begin{equation}
\begin{aligned}
\sigma_y^2&\pqty{\tau}  =
\frac{1}{2\tau^2}\:\vb{E}\bqty{\pqty{\alpha\pqty{t+2\tau}-2\,\alpha\pqty{t+\tau}+\alpha\pqty{t}}^2}\\
&= \frac{1}{2\tau^2}\:\vb{E}\bqty{\pqty{
    \pqty{\alpha\pqty{t+2\tau}-\alpha\pqty{t+\tau}} -
    \pqty{\alpha\pqty{t+\tau}-\alpha\pqty{t}}}^2  }\\
&= \frac{1}{2\tau^2} { \vb{E}\bqty{
    \pqty{\alpha\pqty{t+2\tau}-\alpha\pqty{t+\tau}}^2}} \\ &
\quad\quad\quad\quad + \frac{1}{2\tau^2} {\vb{E}\bqty{ 
    \pqty{\alpha\pqty{t+\tau}-\alpha\pqty{t}}^2  } }\\
&= \frac{1}{2\tau^2} \pqty{c\,\tau + c\,\tau} = \frac{c}{\tau}
\end{aligned}
\label{eqn:allanvaralphawiener}  
\end{equation}

%% file: ms.bbl
\begin{thebibliography}{10}
\providecommand{\url}[1]{#1}
\csname url@samestyle\endcsname
\providecommand{\newblock}{\relax}
\providecommand{\bibinfo}[2]{#2}
\providecommand{\BIBentrySTDinterwordspacing}{\spaceskip=0pt\relax}
\providecommand{\BIBentryALTinterwordstretchfactor}{4}
\providecommand{\BIBentryALTinterwordspacing}{\spaceskip=\fontdimen2\font plus
\BIBentryALTinterwordstretchfactor\fontdimen3\font minus
  \fontdimen4\font\relax}
\providecommand{\BIBforeignlanguage}[2]{{%
\expandafter\ifx\csname l@#1\endcsname\relax
\typeout{** WARNING: IEEEtran.bst: No hyphenation pattern has been}%
\typeout{** loaded for the language `#1'. Using the pattern for}%
\typeout{** the default language instead.}%
\else
\language=\csname l@#1\endcsname
\fi
#2}}
\providecommand{\BIBdecl}{\relax}
\BIBdecl

\bibitem{chaste2012nanomechanical}
J.~Chaste, A.~Eichler, J.~Moser, G.~Ceballos, R.~Rurali, and A.~Bachtold, ``A
  nanomechanical mass sensor with yoctogram resolution,'' \emph{Nature
  Nanotechnology}, vol.~7, no.~5, p. 301, 2012.

\bibitem{naik2009towards}
A.~K. Naik, M.~Hanay, W.~Hiebert, X.~Feng, and M.~L. Roukes, ``Towards
  single-molecule nanomechanical mass spectrometry,'' \emph{Nature
  Nanotechnology}, vol.~4, no.~7, p. 445, 2009.

\bibitem{hanay2012single}
M.~Hanay, S.~Kelber, A.~Naik, D.~Chi, S.~Hentz, E.~Bullard, E.~Colinet,
  L.~Duraffourg, and M.~Roukes, ``Single-protein nanomechanical mass
  spectrometry in real time,'' \emph{Nature Nanotechnology}, vol.~7, no.~9, pp.
  602--608, 2012.

\bibitem{feng2008self}
X.~Feng, C.~White, A.~Hajimiri, and M.~L. Roukes, ``A self-sustaining
  ultrahigh-frequency nanoelectromechanical oscillator,'' \emph{Nature
  Nanotechnology}, vol.~3, no.~6, pp. 342--346, 2008.

\bibitem{van2013nonlinear}
R.~Van~Leeuwen, D.~Karabacak, H.~Van~der Zant, and W.~Venstra, ``Nonlinear
  dynamics of a microelectromechanical oscillator with delayed feedback,''
  \emph{{Physical Rev B}}, vol.~88, no.~21, p. 214301, 2013.

\bibitem{cleland2002noise}
A.~Cleland and M.~Roukes, ``Noise processes in nanomechanical resonators,''
  \emph{Journal of Applied Physics}, vol.~92, no.~5, pp. 2758--2769, 2002.

\bibitem{ekinci2004ultimate}
K.~Ekinci, Y.~Yang, and M.~Roukes, ``Ultimate limits to inertial mass sensing
  based upon nanoelectromechanical systems,'' \emph{{Journal of Applied Physics}},
  vol.~95, no.~5, pp. 2682--2689, 2004.

\bibitem{gavartin2013stabilization}
E.~Gavartin, P.~Verlot, and T.~J. Kippenberg, ``Stabilization of a linear
  nanomechanical oscillator to its thermodynamic limit,'' \emph{Nature
  Communications}, vol.~4, p. 2860, 2013.

\bibitem{Sansa2016}
M.~Sansa, E.~Sage, E.~C. Bullard, M.~G{\'e}ly, T.~Alava, E.~Colinet, A.~K.
  Naik, L.~G. Villanueva, L.~Duraffourg, M.~L. Roukes, G.~Jourdan, and
  S.~Hentz, ``Frequency fluctuations in silicon nanoresonators,'' \emph{Nature
  Nanotechnology}, vol.~11, no.~6, pp. 552--558, June 2016.

\bibitem{kenig2012optimal}
E.~Kenig, M.~Cross, L.~Villanueva, R.~Karabalin, M.~Matheny, R.~Lifshitz, and
  M.~Roukes, ``Optimal operating points of oscillators using nonlinear
  resonators,'' \emph{Physical Rev E}, vol.~86, no.~5, p. 056207, 2012.

\bibitem{villanueva2013surpassing}
L.~Villanueva, E.~Kenig, R.~Karabalin, M.~Matheny, R.~Lifshitz, M.~Cross, and
  M.~Roukes, ``Surpassing fundamental limits of oscillators using nonlinear
  resonators,'' \emph{{Physical Rev Letters}}, vol. 110, no.~17, p. 177208,
  2013.

\bibitem{kenig2013phase}
E.~Kenig, M.~Cross, J.~Moehlis, and K.~Wiesenfeld, ``Phase noise of oscillators
  with unsaturated amplifiers,'' \emph{{Physical Rev E}}, vol.~88, no.~6, p.
  062922, 2013.

\bibitem{demir2018numerical}
A.~Demir and M.~Hanay, ``Numerical analysis of multi-domain systems: Coupled
  nonlinear {PDEs} \& {DAEs} with noise,'' \emph{IEEE Transactions on
  Computer-Aided Design of Integrated Circuits \& Systems}, July 2018.

\bibitem{roy2018improving}
S.~K. Roy, V.~T.~K. Sauer, J.~N. Westwood-Bachman, A.~Venkatasubramanian, and
  W.~K. Hiebert, ``Improving mechanical sensor performance through larger
  damping,'' \emph{Science}, vol. {360, eaar5220}, 2018.

\bibitem{yuksel2019nonlinear}
M.~Yuksel, E.~Orhan, C.~Yanik, A.~B. Ari, A.~Demir, and M.~S. Hanay,
  ``Nonlinear nanomechanical mass spectrometry at the single-nanoparticle
  level,'' \emph{arXiv:1902.02520 preprint}, 2019.

\bibitem{kelleci2018towards}
M.~Kelleci, H.~Aydogmus, L.~Aslanbas, S.~O. Erbil, and M.~S. Hanay, ``Towards
  microwave imaging of cells,'' \emph{Lab on a Chip}, vol.~18, no.~3, pp.
  463--472, 2018.

\bibitem{hauer2013general}
B.~Hauer, C.~Doolin, K.~Beach, and J.~Davis, ``A general procedure for
  thermomechanical calibration of nano/micro-mechanical resonators,''
  \emph{Annals of Physics}, vol. 339, pp. 181--207, 2013.

\bibitem{olcum2015high}
S.~Olcum, N.~Cermak, S.~C. Wasserman, and S.~R. Manalis, ``High-speed
  multiple-mode mass-sensing resolves dynamic nanoscale mass distributions,''
  \emph{Nature Communications}, vol.~6, p. 7070, 2015.

\bibitem{benedetto1999principles}
S.~Benedetto and E.~Biglieri, \emph{Principles of digital transmission: with
  wireless applications}.\hskip 1em plus 0.5em minus 0.4em\relax Springer
  Science \& Business Media, 1999.

\bibitem{yurke1995theory}
B.~Yurke, D.~Greywall, A.~Pargellis, and P.~Busch, ``Theory of amplifier-noise
  evasion in an oscillator employing a nonlinear resonator,'' \emph{Physical
  Rev A}, vol.~51, no.~5, p. 4211, 1995.

\bibitem{GardnerRP}
W.~Gardner, \emph{Introduction to Random Processes}.\hskip 1em plus 0.5em minus
  0.4em\relax McGraw-Hill, 1990.

\bibitem{RoLoFe97jssc}
J.~Roychowdhury, D.~Long, and P.~Feldmann, ``Cyclostationary noise analysis of
  large {RF} circuits with multi-tone excitations,'' \emph{IEEE Journal of
  Solid-State Circuits}, April 1998.

\bibitem{demir1998Kluwer}
A.~Demir and A.~Sangiovanni-Vincentelli, \emph{Analysis and simulation of noise
  in nonlinear electronic circuits and systems}.\hskip 1em plus 0.5em minus
  0.4em\relax Springer Science \& Business Media, 1998.

\bibitem{allan1966statistics}
D.~W. Allan, ``Statistics of atomic frequency standards,'' \emph{Proceedings of
  the IEEE}, vol.~54, no.~2, pp. 221--230, 1966.

\bibitem{rubiola2009phase}
E.~Rubiola, \emph{Phase noise and frequency stability in oscillators}.\hskip
  1em plus 0.5em minus 0.4em\relax Cambridge University Press, 2009.

\bibitem{wiki:Allanvariance}
Wikipedia, ``Allan variance -- {Wikipedia}{,} {The Free Encyclopedia},'' 2019.

\bibitem{welch_use_1967}
P.~D. Welch, ``The {use} of {Fast} {Fourier} {Transform} for the {estimation}
  of {power} {spectra}: {A} {method} {based} on {time} {averaging} {over}
  {short}, {modified} {periodograms},'' \emph{IEEE Transactions on Audio and
  Electroacoustics}, vol.~15, no.~2, pp. 70--73, June 1967.

\bibitem{karr1953effective}
P.~R. Karr, ``Effective circuit bandwidth for noise with a power-law
  spectrum,'' \emph{Journal of Research of the National Bureau of Standards},
  vol.~5, no.~1, 1953.

\bibitem{Grimmet92}
G.~Grimmet and D.~Stirzaker, \emph{Probability and Random Processes},
  2nd~ed.\hskip 1em plus 0.5em minus 0.4em\relax Oxford Science Publications,
  1992.

\end{thebibliography}
